\documentclass[journal]{IEEEtran}

\usepackage{xcolor,soul,framed} 

\colorlet{shadecolor}{yellow}
\usepackage[pdftex]{graphicx}
\graphicspath{{../pdf/}{../jpeg/}}
\DeclareGraphicsExtensions{.pdf,.jpeg,.png}

\usepackage[cmex10]{amsmath}
\usepackage{array}
\usepackage{bbm}
\usepackage{amssymb}
\usepackage{mdwmath}
\usepackage{mdwtab}
\usepackage{eqparbox}
\usepackage{url}
\usepackage[ruled,linesnumbered]{algorithm2e}
\usepackage{graphicx}
\usepackage{subfig}
\usepackage{multirow}
\usepackage{adjustbox}
\usepackage{dutchcal}
\usepackage{dblfloatfix}
\usepackage{adjustbox}
\usepackage[colorlinks=true,allcolors=blue]{hyperref}
\usepackage{soul, color, xcolor}
\soulregister\cite7
\soulregister\ref7 
\usepackage{tcolorbox}
\usepackage[flushleft]{threeparttable}
\usepackage{xcolor}
\usepackage{orcidlink}

\begin{document}

    \title{Real-time and Downtime-tolerant Fault Diagnosis for Railway Turnout Machines (RTMs) Empowered with Cloud-Edge Pipeline Parallelism}

\author{Fan Wu~\orcidlink{0000-0002-2787-7860},
        Muhammad Bilal~\orcidlink{0000-0003-4221-0877},~\IEEEmembership{Senior Member,~IEEE,}
        Haolong Xiang,
        Heng Wang,
        Jinjun Yu,
        and~Xiaolong Xu,~\IEEEmembership{Senior Member,~IEEE}
\thanks{Manuscript received 26 March 2024; revised 6 November 2024; accepted 9 February 2024. This work was supported in part by the National Natural
Science Foundation of China under Grant 62372242 and Grant 92267104 and in part by the Natural Science Foundation of Jiangsu Province of China under
Grant BK20211284. (Corresponding authors: Xiaolong Xu.)}%
\thanks{Fan Wu is with the School of Software, Nanjing University of Information Science and Technology, Nanjing 210044, China (e-mail: zzxjl1@hotmail.com).}
\thanks{Muhammad Bilal is with the School of Computing and Communications, Lancaster University, Bailrigg, LA1 4WA Lancaster, U.K. (e-mail: m.bilal@ieee.org).}%
\thanks{Haolong Xiang is with the School of Software, Nanjing University of Information Science and Technology, Nanjing 210044, China (e-mail: hlx6700@gmail.com)}%
\thanks{Heng Wang is with the NRIET Industrial Co.,Ltd., Nanjing 211106, China (e-mail: wangheng@glarun.com).}%
\thanks{Jinjun Yu is with the NRIET Industrial Co.,Ltd., Nanjing 211106, China (e-mail: yujinjun@glarun.com).}%
\thanks{Xiaolong Xu is with the School of Software, Nanjing University of Information Science and Technology, Nanjing 210044, China (e-mail: xlxu@ieee.org)}%
}

\markboth{}{Wu \MakeLowercase{\textit{et al.}}: Distributed Fault Diagnosis for Railway Turnout Machines}

\maketitle

\begin{abstract} 

Railway Turnout Machines (RTMs) are mission-critical components of the railway transportation infrastructure, responsible for directing trains onto desired tracks. Due to frequent operations and exposure to harsh environments, RTMs are susceptible to failures and can potentially pose significant safety hazards. 
For safety assurance applications, especially in early-warning scenarios, RTM faults are expected to be detected as early as possible on a continuous 7x24 basis.
However, limited emphasis has been placed on distributed model inference frameworks that can meet the inference latency and reliability requirements of such mission-critical fault diagnosis systems, as well as the adaptation of diagnosis models within distributed architectures. This has hindered the practical application of current AI-driven RTM monitoring solutions in industrial settings, where single points of failure can render the entire service unavailable due to standalone deployment, and inference time can exceed acceptable limits when dealing with complex models or high data volumes.
In this paper, an edge-cloud collaborative early-warning system is proposed to enable real-time and downtime-tolerant fault diagnosis of RTMs, providing a new paradigm for the deployment of models in safety-critical scenarios. Firstly, a modular fault diagnosis model is designed specifically for distributed deployment, which utilizes a hierarchical architecture consisting of the prior knowledge module, subordinate classifiers, and a fusion layer for enhanced accuracy and parallelism. Then, a cloud-edge collaborative framework leveraging pipeline parallelism, namely CEC-PA, is developed to minimize the overhead resulting from distributed task execution and context exchange by strategically partitioning and offloading model components across cloud and edge. 
Additionally, an election consensus mechanism is implemented within CEC-PA to ensure system robustness during coordinator node downtime.
Comparative experiments and ablation studies are conducted to validate the effectiveness of the proposed distributed fault diagnosis approach. Our ensemble-based fault diagnosis model achieves a remarkable 97.4\% accuracy on a real-world dataset collected by Nanjing Metro in Jiangsu Province, China. Meanwhile, CEC-PA demonstrates superior recovery proficiency during node disruptions and speed-up ranging from 1.98x to 7.93x in total inference time compared to its counterparts.

\end{abstract}


%
\IEEEpeerreviewmaketitle


\section{Introduction} \label{Intro}

\IEEEPARstart{R}{ailway} transportation offers a high-capacity, cost-effective, and environmentally friendly solution for long-distance travel, making it a popular choice for passenger and freight services in Europe, Asia, and North America. According to M\&M market research \cite{railway_market_report}, the global railway system was valued at \$25.1 billion in 2022 and is estimated to reach \$30.9 billion by 2027. The Railway Turnout Machines (RTMs), also known as the Railway Point Machines (RPMs), are critical components of the railway transportation infrastructure, responsible for directing trains onto desired tracks. However, RTMs are prone to failures due to wearing caused by frequent operations and exposure to harsh outdoor environments. Statistical analysis reveals RTMs as one of railside equipment that experience the highest failure rates, accounting for 18\% of all documented railway system failures occurring between 2011 and 2017 \cite{GROSSONI2021104987}. The malfunction of RTMs can lead to catastrophic accidents such as collisions and train derailments, resulting in severe casualties and property losses. This typically involves the concept of preventive maintenance \cite{WOS:001068130400001}, which calls for regularly scheduled inspections and repairs targeting at the prevention of failures before they occur. For a long time, such condition-based maintenance mainly depends on the expert knowledge and experience of railway workers and thus can be time-consuming and labor-intensive. Therefore, an unsupervised, resilient, and responsive RTM fault early-warning system for train drivers and maintenance groups has raised lots of concern in the industry.

With the advent of information technology, Railside Monitoring Units (RMUs) are deployed to collect runtime data during the operation of RTMs. Numerous fault diagnosis methods have been developed utilizing the collected data on vibration \cite{9786783}, current \cite{doi:10.1177/0361198119837222,10038650,9104763}, torque and acoustic signals \cite{9531564}, etc. Previous endeavors have been primarily dedicated to enhancing model accuracy, while paying little attention to the performance and reliability issues caused by inappropriate deployment methods \cite{9745085}. For safety assurance applications, especially in early-warning scenarios, we expect faults to be detected as early as possible to provide drivers and maintenance groups with more response time. The high computational overhead and complex procedures of these fault diagnosis models can make real-time inference challenging on resource-constrained devices such as Personal Computers (PCs). The traditional standalone deployment \cite{9579018}, where all the model components are deployed on a single device or platform, is also susceptible to system-wide unavailability in case of any software or hardware malfunctions on that centralized node \cite{10.5555/2643634.2643666}.

Cloud computing has then become a common approach to wide range of fault diagnostic applications in Industry 4.0 \cite{WOS:001129667400003}, micro-electromechanical systems (MEMS) \cite{10026418}, Cloud Native \cite{WOS:001306799100041}, etc. However, the data gathered must be sent to the cloud to harness the high-performance and elastic advantages of cloud computing. In addition to privacy concerns \cite{WOS:000965227300001} stemming from the sensitive nature of sensor data (e.g., route schedules and geographical locations), the transmission of data in railway environments like underground tunnels, inevitably leads to data loss and network latency issues \cite{gheth2021communication}. These factors significantly impair the real-time capabilities of cloud-based solutions and hinder their effectiveness in monitoring mission-critical infrastructure \cite{guo2021towards}.

In the past decade, academic interest has grown in combining edge computing with fault detection for model deployment, also known as Edge Intelligence (EI) \cite{gong2023edge}. This novel approach shifts computation from centralized cloud servers to the network edge, offering latency \cite{WOS:000965831400001}, energy consumption \cite{WOS:000740006200055}, Quality of Service (QoS) \cite{WOS:000808096100085} and mobility \cite{WOS:000745475500001} enhanced solutions. Federated Learning (FL) \cite{9800698} has emerged as a potent approach for preserving privacy during model training, which enable each distributed client to train a local replica of the global model with its own dataset before sending updates to aggregate the shared global model.

However, limited emphasis has been placed on distributed model inference frameworks that can meet the latency and reliability requirements of the fault diagnosis model deployment, or on tailoring the diagnosis models to perform optimally within distributed architectures. 
The inherent complementarity of cloud and edge computing has fostered the concept of cloud-edge collaboration \cite{9681206}, a paradigm that dynamically allocates and coordinates computational tasks across cloud and edge. This collaborative approach has inspired new paradigms for AI-driven real-time and downtime-tolerant monitoring tasks in mission-critical industrial applications \cite{WOS:001258244000085} , where such systems benefit from the high availability characteristic of modern cloud computing infrastructure and the low-latency capabilities afforded by edge computing deployments.
Therefore, a RTM fault diagnosis model optimized for distributed deployment, coupled with its edge-cloud collaboration empowered model inference framework is proposed in this paper, where model components are strategically partitioned and offloaded jointly across cloud and edge rather than relying solely on cloud or local to facilitate reliability and faster response. 

The main contributions of this paper can be summarized as:
\begin{itemize}
    \item A parallel-optimized RTM fault diagnosis model is developed with model integration technique. The model incorporates an enhanced three-stage segmentation scheme as prior knowledge and the outputs of multiple sub-classifiers are fused by a fuzzy-based ensemble mechanism to form the final classification result.
    \item A cloud-edge collaborative framework leveraging pipeline parallelism, namely CEC-PA, is proposed to address the real-time and robustness challenges of distributed fault diagnosis. CEC-PA partitions the integrated model components into pipelines and intelligently schedules them across all worker nodes. Additionally, a downtime-tolerant mechanism is proposed to ensure system robustness.
    \item Extensive experiments are conduced to evaluate the effectiveness of the proposed fault detection model and CEC-PA framework. Results showcase our ensemble-based fault diagnosis model produce accurate predictions across all fault types and CEC-PA outperform other approaches in terms of real-time performance and reliability.
\end{itemize}

The rest of this paper is organized as follows: Section \ref{related_work} discusses previous works on parallelization techniques in distributed AI. Section \ref{preliminary} presents the preliminary discussion on the working principle and current pattern analysis of three-stage turnouts. Section \ref{model_formulation} establishes the time consumption model and multi-objective optimization problem of the proposed cloud-edge RTM fault early-warning system. Section \ref{fault_diag} implements the parallel-optimized turnout fault diagnosis scheme and provides a detailed description of the interactions between each module. Section \ref{section_cec} presents the design details of CEC-PA. Section \ref{experiments} demonstrates the effectiveness of the fault diagnosis model and CEC-PA through comparative experiments. Finally, Section \ref{conclusion} draws a conclusion of this paper and highlights its future research directions.

\begin{table*}[b]
\renewcommand{\arraystretch}{1.5}
\centering
\caption{Comparison of Different Parallelization Strategies}
\begin{tabular}{cccc}
\hline
\textbf{Key Characteristics} & \textbf{Data Parallelization} & \textbf{Model Parallelization} & \textbf{Pipeline Parallelization} \\ 
\hline
Applicable scenarios & Large datasets with smaller models & Extremely large models & Long pipelines\\
Proof of convergence & $\checkmark$ & $\times$ & $\checkmark$ \\
Heterogeneous cluster support & $\checkmark$ & $\times$ & $\checkmark$ \\
Load balance & $\times$ & $\checkmark$ & $\checkmark$ \\
Communication overhead & High & Low & Moderate\\
Implementation difficulty & Low & High & Moderate \\
Scalability & High & Moderate & High\\
\hline
\end{tabular}
\label{tab:parallelization}
\end{table*}

\section{Related Work} \label{related_work}

\subsection{Intelligent Health Monitoring for RTMs}
Numerous solutions for unmanned intelligent RTM health monitoring have been proposed in the past two decades. 
Ou et al. \cite{doi:10.1177/0361198119837222} proposed a RTM fault diagnosis scheme based on Machine Learning (ML), where a modified Support Vector Machine (SVM) with Gaussian kernel is applied to classify the time-domain and frequency-domain features obtained through Linear Discriminant Analysis (LDA). Ji et al. \cite{9956825} introduce an adaptive fault diagnosis model for both single and double-action RTMs that utilizes Dynamic Time Warping (DTW) to calculate similarities between input samples and their built-in reference templates. Wang et al. \cite{10038650} leverage segmentalized Max-Relevance and Min-Redundancy (mRMR) techniques for stage-wise feature extraction. Additionally, a novel classifier named cost-sensitive Extreme Learning Machine (cf-ELM) is complemented in their study, which features bias compensation to enhance classification stability. Deep learning (DL) approaches are also widely adopted due to their strong generalization capabilities. By creating variants of Deep Auto Encoders (DAEs) and Gated Recurrent Units (GRUs), Zhang et al. \cite{9579018} and Guo et al. \cite{9104763} propose adaptive latent feature classification method for unsupervised and semi-supervised RTM fault diagnosis, respectively. 

Other than using common data inputs such as power spectral density and current sequence, Cao et al. provide a distinct perspective by focusing on alternative fault diagnosis methodologies leveraging acoustic data \cite{9531564} and three-dimensional vibration signals \cite{9786783}. Additionally, to address the Few-shot Fault Diagnosis (FSFD) problem where limited faulty samples are available, Li et al. \cite{10003128} developed a reweighted regularized prototypical network combined with a novel balance-enforcing regularization (BER) mechanism to hedge against the between-class imbalance and improve classification accuracy.

\subsection{Parallelization Techniques in Distributed AI}
According to Mwase et al. \cite{MWASE2022292}, parallelism in Distributed AI can be carried out by breaking down either the data, the model, the stages of the process (i.e., pipeline), or a combination of these. 
Table \ref{tab:parallelization} presents the key characteristics of these well-established parallelization techniques.

Data parallelization emerges as a highly effective strategy for accelerating DL on Graphics Processing Units (GPUs), offering versatility and ease of implementation. In this approach, the input batch of dataset is spilt into multiple micro-batches, each allocated to a distinct data-parallel worker. Pandey et al. \cite{WOS:000772442700003} experimentally demonstrated that the implementation of data parallelization at small scales can achieve near-perfect scaling due to the combination of independent computations and low computational density. Foundation models such as GPT and SAM have demonstrated state-of-the-art performance on various tasks in Natural Language Processing (NLP) and Computer Vision (CV). As a result, such heavyweight models (GPT-3 typically with 175 billion parameters) are too large to fit on a single device and if so still take forever to train. 

Two parallelization techniques have emerged to mitigate these issues: model parallelism and pipeline parallelism. In contrast to data parallelism, model parallelism (i.e., tensor parallelism)  addresses storage limitations via model partitioning and minimizes communication overhead by avoiding complete parameter transfers during each update iteration \cite{MWASE2022292}. Leveraging model parallelism techniques, Xu et al. \cite{10177476} and Lai et al. \cite{WOS:000954366000004} proposed SUMMA and DeCNN for the efficient and scalable training of large-scale DL models. However, Gomez et al. \cite{10.5555/3586589.3586760} point out that model parallelism places extremely high demands on low-latency and high-throughput interconnection between GPUs. Therefore, its usage is restricted by proprietary hardware, such as NVLink, and thus limits the potential for pipeline parallelism to be widely deployed on diverse computing platforms. After analyzing the communication overhead of different parallelization strategies, Oyama et al. \cite{WOS:000621405200012} concluded that pipeline parallelism divides the layers of the model into stages that only shares activations between neighboring pipeline stages, resulting in even lower communication overhead compared with its model parallelism predecessor. In the context of affordable training of large DNNs, Thorpe et al. introduced Bamboo \cite{285072}, a distributed system that introduces redundant computations into the training pipeline to provide resilience at a low cost, outperforming traditional checkpointing in training throughput and reducing costs. Additionally, Zhao et al. \cite{WOS:000683977300001} and Kim et al. \cite{pmlr-v202-kim23l} demonstrate that pipeline parallelism can accelerate processing without any accuracy loss, as opposed to compression techniques like pruning and quantization.

\section{Preliminary} \label{preliminary}

\subsection{Turnout Fault Diagnosis via Current Monitoring}

Turnout machines can be classified into three categories: electro-hydraulic, electro-mechanical, and all-electric \cite{GROSSONI2021104987}. In this paper, we will focus on the most commonly used electro-mechanical modules, which consist of major components including the electric motor, mechanical parts (gear box, friction clamp, locking rod, etc.), and control circuits. Based on its electrical characteristics, we denote the input voltage as $U$, the input current as $I$, the three-phase angle as $\theta$, motor’s angular velocity as $\Omega$, and efficiency as $\eta$. During normal operation, the correlation between power $P$ of the motor and its output torque $T$ can be expressed as

\small{
\begin{equation}
\label{RTM_TORQUE_POWER}
T = \frac{P}{\Omega} = \frac{\sqrt{3} \eta U I \cos \theta}{\Omega}.
\end{equation}
}

As the sampling interval of MMS (Microcomputer Monitoring System) is typically small (usually less than 100ms), $\Omega$ can be approximated as constant over this duration. According to Equation (\ref{RTM_TORQUE_POWER}), torque $T$ is positively correlated with motor power $P$ and current $I$. Any variations in resistive forces acting on the motor shaft during switching, such as control system state transitions, mechanical obstructions, or lubrication deficiencies, will manifest as fluctuations in the current waveform $I$. Therefore, real-time monitoring and analysis of the motor current can provide insights into the working status of the turnout module.

\subsection{Current Pattern Analysis of Three-Stage Turnouts}

\begin{figure}[h]
  \begin{center}
  \includegraphics[width=0.75\linewidth]{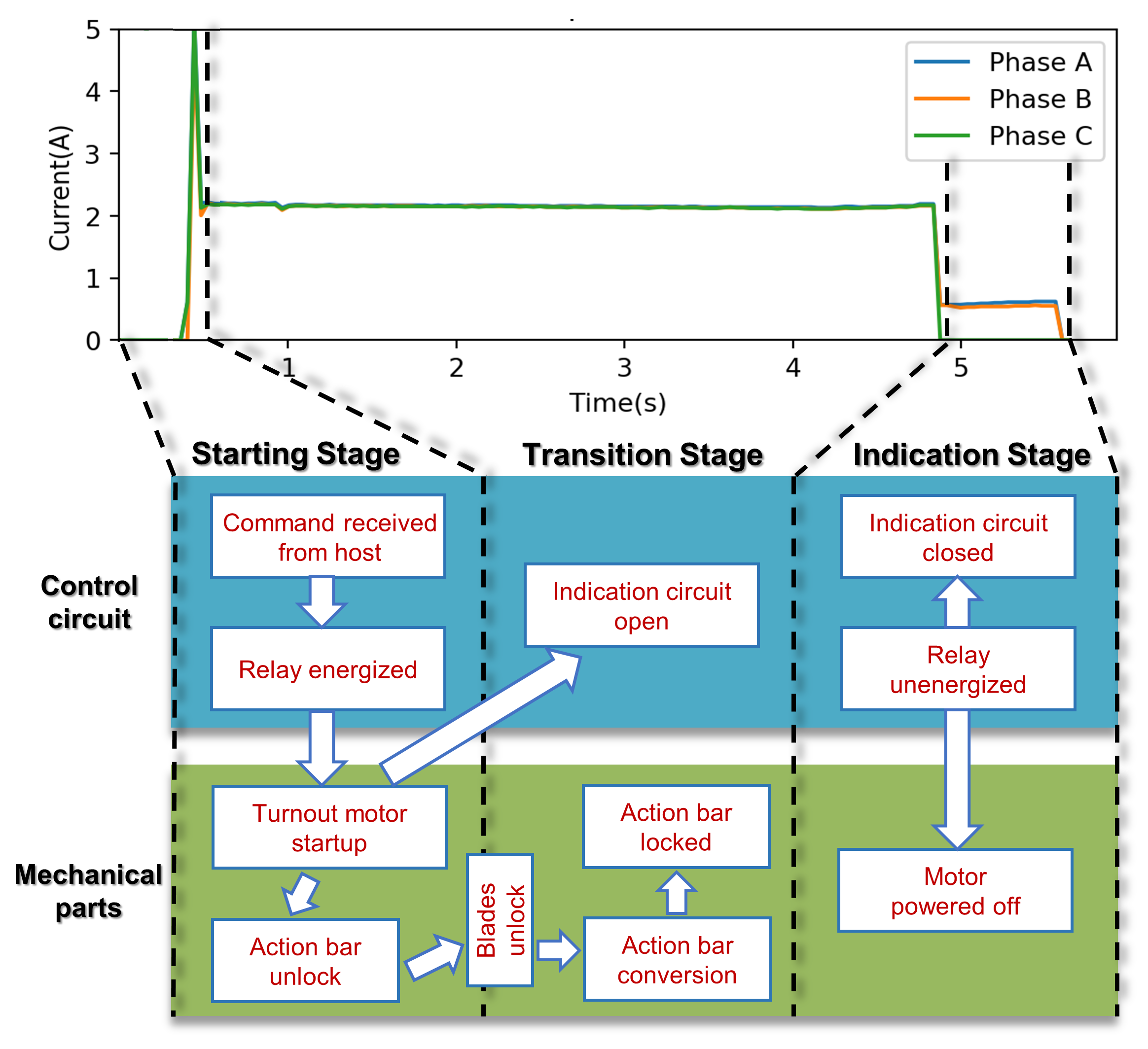}
  \caption{Decomposed analysis of RTM current sequence.}
  \label{RTM_CURRENT_DECOMP}
  \end{center}
  \vspace{-2em}
\end{figure}

According to \cite{9579018,doi:10.1177/0361198119837222}, the current waveform during turnout transitions (i.e., from normal to reverse position and vice versa) follows a characteristic three-stage profile that closely matches the module's operational procedure. As illustrated in Figure \ref{RTM_CURRENT_DECOMP}, these stages can be outlined as follows:

\textbf{a) Starting Stage:} Initially, all three phase currents are zero as the control circuit relay only energizes after a built-in time delay. Upon motor startup, a large current surge rising from 0 occurs due to efforts overcome rotor inertia and the unlocking resistance between the stock rails and closure rails. However, once the motor reaches its operating speed, the current will decline to a relatively constant level.

\textbf{b) Transition Stage:} As the motor persists in rotating the drive shaft, it engages the rack mechanism, which facilitates the lateral movement of the switch rails until they are securely locked in place. Throughout this stage, the three-phase current remains steady without any sudden fluctuations or overcurrent conditions.

\textbf{c) Indication Stage:} When RTM has reached its fully locked position, the control circuit relay de-energizes the contactor, disconnecting the motor terminals. This causes the current in one phase to rapidly decrease to zero. However, owing to the RTM's buffering effect, the other two current phases maintain a constant value of approximately 0.6 Amp before eventually dropping to zero.

\section{Model Formulation and Problem Definition} \label{model_formulation}
\subsection{Network Topology}	
\label{section_4.1}

\begin{figure*}[]
\vspace{-1em}
  \begin{center}
  \includegraphics[width=0.65\linewidth]{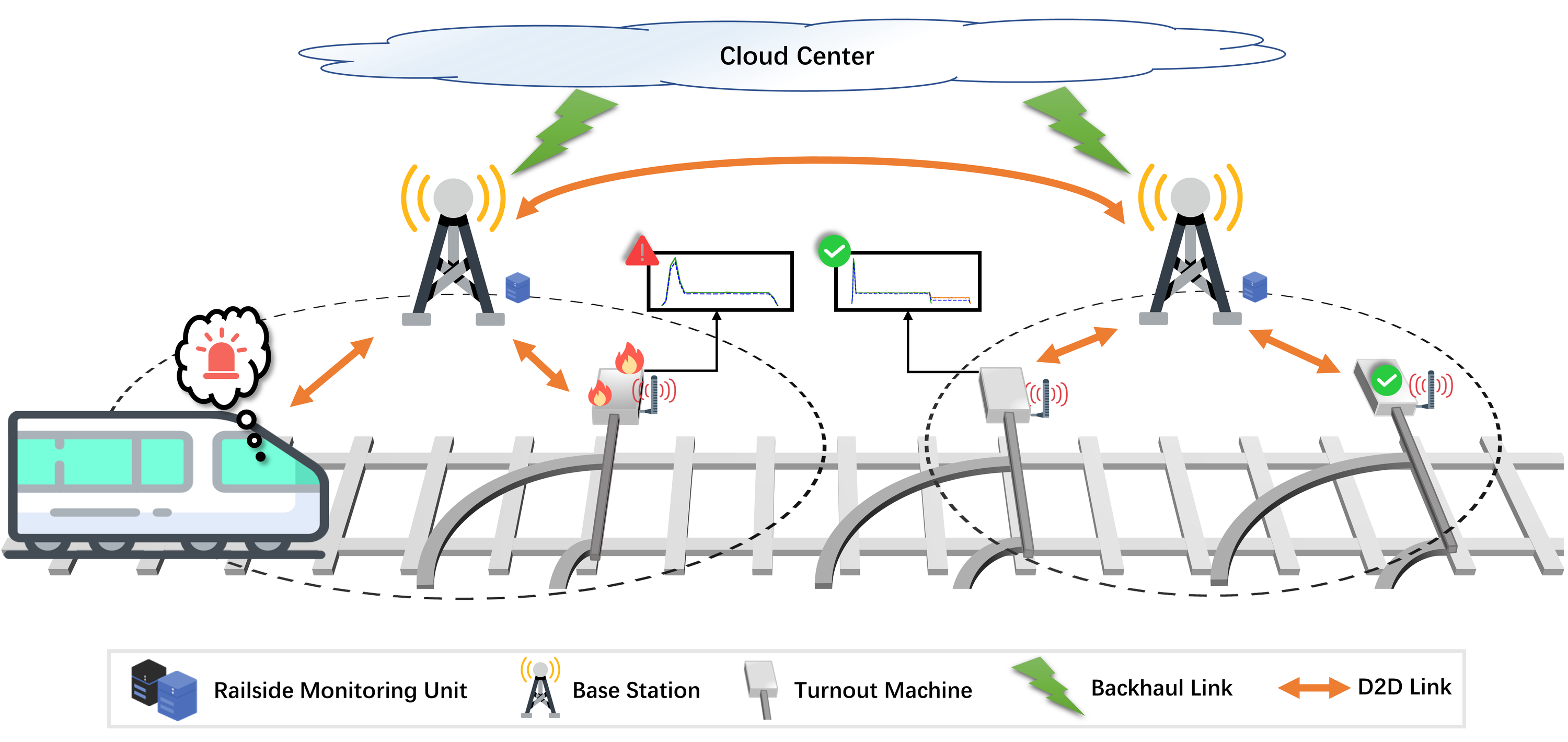}
  \caption{Network topology of the proposed track anomaly early-warning system.}
  \label{network_topology}
  \end{center}
\vspace{-2em}
\end{figure*}

High-speed trains require extensive safety precautions to prevent accidents due to track irregularities. This paper presents a track anomaly early-warning system consisting of high-speed trains, cloud center, RMUs, Base Stations (BSs), and turnout machines. A heterogeneous network paradigm is employed to establish interconnection between these components, as depicted in Figure \ref{network_topology}.

Turnout machines $M = \{m_1, m_2, \ldots, m_S\}$ are all equipped with MMS current sensing modules to continuously sample operational data during each duty cycle. RMUs $R = \{r_1, r_2, \ldots, r_J\}$ are strategically distributed along the tracks at regular intervals to collect data from adjacent turnout machines. In addition to aggregating sensory outputs into built-in storage, the RMUs possess moderate on-board computation abilities to serve the purpose of executing computation offloading instructions. Each $r_j$ is assigned to a dedicated BS, which is responsible for transmitting and receiving data within coverage range $g_j$. The cloud center $C$ is capable of high-concurrency task execution while also responsible for task scheduling decisions. Short-range wireless device-to-device (D2D) communication is established between RTMs and BSs leveraging the full-duplex IEEE 802.11p protocol \cite{WOS:000510677500008}. High-speed trains, denoted as $V = \{v_1, v_2, \ldots, v_K\}$, are also equipped with onboard electronics to directly communicate with BSs in a D2D manner as they traverse along the tracks. Additionally, in the event of cloud center failures or defective backhaul connections, backup connections between adjacent RMUs can be established to form a self-organized mesh network for uninterrupted data transmission.

The proposed early-warning system employs a three-tier architecture where high-speed trains $V$, RMUs $R$, and the cloud center $C$ function as end devices, edge nodes, and the cloud, respectively. Fault classification model inference is collaboratively executed across $R$ and $C$. Upon collecting RTM data, $R$ initiate resource scheduling requests to $C$. The cloud $C$ subsequently optimizes pipeline partitioning and offloading strategies based on current node status and network congestion. Assigned tasks and parameters are then distributed to $R$ for execution, with results subsequently aggregated in $C$. Following fault diagnosis model inference completion, detected anomalies trigger network-wide broadcasts. Throughout operations, $V$ maintain continuous D2D communication with $R$ in-range, receiving real-time diagnostic information and responding to fault warnings as necessary.

\subsection{Distributed Task Execution Model}
\label{section_4.2}
Traditional fault diagnosis systems \cite{9956825, 9579018} typically adopt a centralized approach where data is processed on a single node. This study designs a parallel distributed model where the overall workflow is partitioned into discrete subtasks that can be executed concurrently across edge and cloud. Let $CT(t) = \{ct_1, ct_2, \ldots, ct_n\}$ denote the set of computing tasks generated at time slot $t$. Each task $ct_i$ can be further decomposed into a sequence of fine-grained subtasks $\{\tau_1, \tau_2, \ldots, \tau_{N(ct_i)}\}$, where $N(ct_i)$ represents the number of subtasks decomposed from $ct_i$. According to Section \ref{section_4.1}, the network consists of $J$ edge node workers $\{W_{r_1}, W_{r_2}, \ldots, W_{r_J}\}$ and one cloud worker $W_C$. Subtasks from the overall task pool $CT(t)$ are scheduled adaptively to workers for processing, ensuring that each worker $W_j$ is assigned a subset of tasks $P_j \subseteq CT(t)$. After completing $P_j$, worker $W_j$ transmits intermediate outputs to the subsequent workers along the workflow, until reaching the High-speed Trains $V$. Assuming there is no data dependency between subtasks $\tau_2$ and $\tau_3$, these independent subtasks are therefore dynamically scheduled in parallel across multiple workers. In contrast, we assume the outputs of $\tau_2$ and $\tau_3$ must be combined to form the required input for $\tau_4$. This introduces an inter-subtask dependency scenario, whereby worker $W_{\leftarrow \tau_4}$ assigned with $\tau_4$ can only proceed once the predecessors $\tau_2$ and $\tau_3$ are completed.

To efficiently schedule and monitor distributed subtasks, we define the characteristics of subtask $\tau_n$ as a quadruple $\tau_n = \langle \varsigma, \Psi, \zeta, \Phi \rangle$, where $\varsigma$, $\Psi$, $\zeta$, and $\Phi$ represent the estimated computational workload, minimal system requirement, predecessor tasks set, and current state quantity, respectively. The life cycle of a successful task involves several key stages, and the computational duration for each stage can be calculated as follows:

\textbf{a) Queueing Stage:} The waiting period from task submission to processing start. Due to the limited hardware capacity and context switching overhead, RMU workers $W_{i \in R}$ can only process a finite number of tasks concurrently. Queue $Q = \{\tau_1, \tau_2, \ldots\}$ is used to store tasks that cannot be immediately processed. Workers are strategically selected from the set $\mathbb{A} = \{W_i \mid sys\_res(r_j) \geq \Psi\}$ to maximize overall system performance. Subsequently, subtasks are added to the corresponding $W_i$’s queue, awaiting execution until concurrency limit $\omega$ is no longer exceeded. The queueing time of $\tau_n$ scheduled for execution on worker $W_i$ can then be modeled as

\small{
\begin{equation}
t_{W_i}^{{queue}}(\tau_n) = \begin{cases} 
0, & {concurr} \leq \omega \\
\sum_{k=1}^{{len}(Q_i)} {duration}(\tau_k), & {concurr} > \omega
\end{cases},
\end{equation}
}

\noindent where ${duration}(\tau_k)$ refers to the estimated processing time of the $k$-th task in the queue, and $Q_i$ represents the queue of worker $W_{r_i}$.

Due to the uncertainty in task completion time and to enhance the system robustness, tasks can be competitively queued across multiple workers. Let $\Gamma$ denote the time required for decision model inference. The total time is given by the minimum time taken among all workers, expressed as

\small{
\begin{equation}
T_{{QUEUE}}(\tau_n) = \Gamma + \min_{W_i \in \mathbb{A}} (t_{W_i}^{{queue}}(\tau_n)).
\end{equation}
}

\textbf{b) Task Execution Stage:} Inspired by \cite{9681206}, Floating-Point Operations (FLOPs) are employed to quantify the computational workload of tasks. FLOPs provide a hardware-agnostic metric for computational workload. For a given CPU with clock frequency of $\phi$ Hz, the number of FLOPs executed per second can be calculated as

\small{
\begin{equation}
\epsilon = \frac{\rho \cdot \chi}{\phi},
\end{equation}
}

\noindent where $\rho$ indicates instruction-level parallelism (i.e., operations per instruction), which captures the pipeline efficiency of the processor architecture. Besides, $\chi$ is the number of Instructions Per Clock (IPC).

Modern processors feature multi-core designs, integrating two or more independent cores within a single chip. This allows independent tasks to run in true parallel, thereby enhancing overall throughput. The total execution time of subtask $\tau_n$ executed on a processor with $N_{cores}$ can be modeled as

\small{
\begin{equation}
T_{{COMP}}(\tau_n) = \frac{\omega}{N_{cores}} \cdot \frac{\varsigma}{\epsilon} = \frac{\phi \omega \varsigma}{N_{cores} \rho \chi}.
\end{equation}
}

\textbf{c) Idle Suspended Stage:} In a distributed workflow, certain subtasks may have dependencies on outputs from their predecessors, causing them to enter an idle suspended state until these dependencies are resolved. We assume $\tau_n$ is a subtask that directly depends on predecessor subtasks $\mathbb{D} = \{\tau_m, \tau_{m+1}, \ldots, \tau_{n-1}\}$. The waiting time of $\tau_n$ for task suspension, also known as the idle time, can be expressed as

\small{
\begin{equation}
\begin{aligned}
T_{{IDLE}}(\tau_n) &= \max\{T_{{COMP}}(\tau_m), \ldots, T_{{COMP}}(\tau_{n-1})\}, \\
& \quad \forall \Phi_k \neq 0, m \leq k \leq n-1,
\end{aligned}
\end{equation}
}

\noindent where $\Phi_k$ is the state quantity of subtask $\tau_k$, and $\Phi_k \neq 0$ indicates $\tau_k$ is still in progress.

By aggregating the contributions of individual subtasks, the total End-to-End execution time of a distributed workflow $ct_i$ can then be modeled as

\small{
\begin{equation}
\begin{aligned}
T_{{EXEC}}(ct_i) &= L + \sum_{n=1}^{N(ct_i)} (1 - l_{{cloud}}(\tau_n)) \cdot T_{{QUEUE}}(\tau_n) \\
& \quad + T_{{COMP}}(\tau_n) + T_{{IDLE}}(\tau_n),
\end{aligned}
\end{equation}
}

\noindent where $\mathbbm{l}_{{cloud}}$ is a Boolean variable indicating if $\tau_n$ is executed on the cloud (value of 1) or edge (value of 0). $L$ represents additional overheads such as failure recovery, connection and state maintenance costs, I/O operations, and Operating System (OS) level expenditures.

\begin{figure*}[t!]
  \begin{center}
  \includegraphics[width=0.85\linewidth]{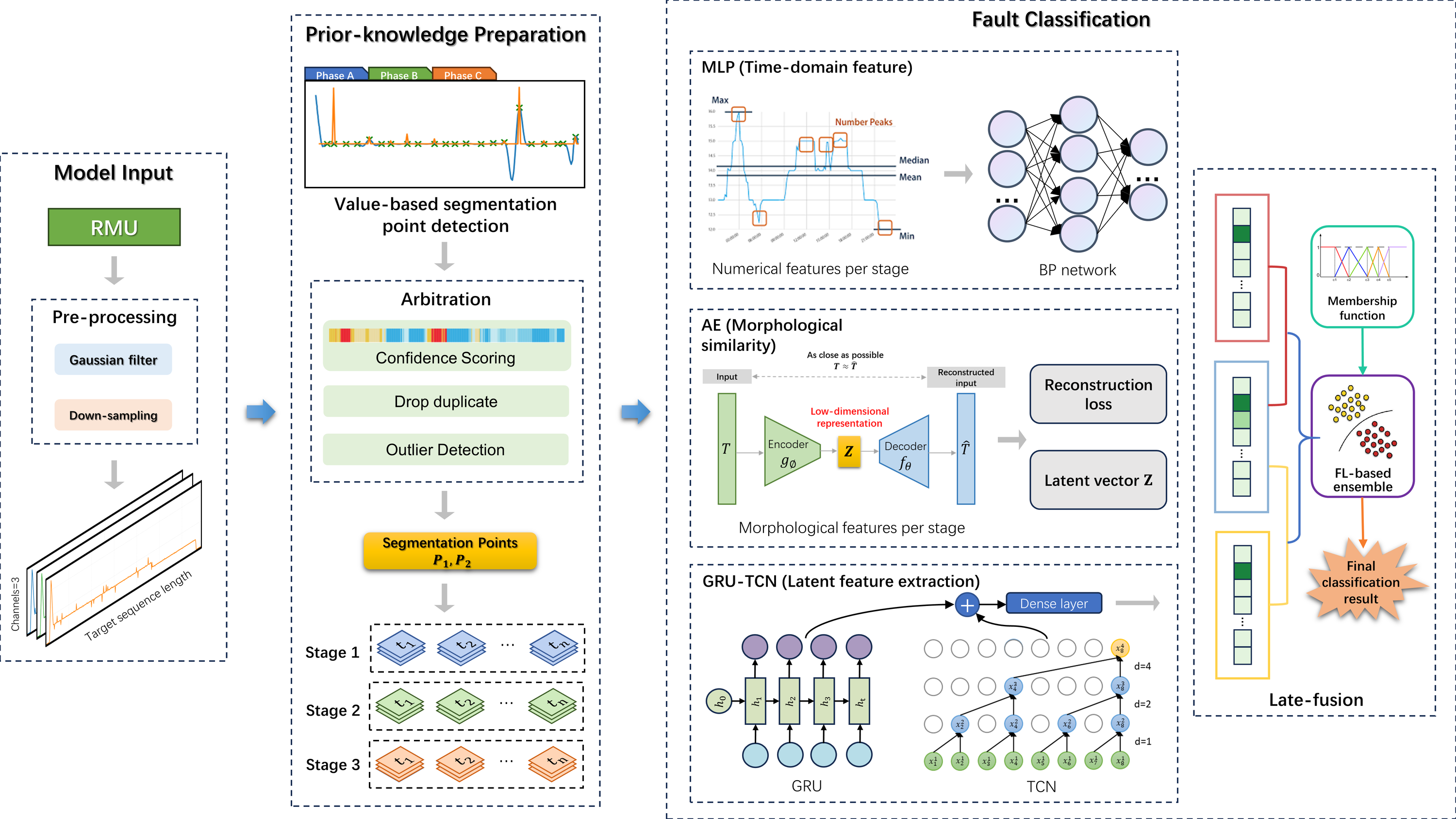}
  \caption{Architecture of the proposed turnout fault diagnosis model leveraging prior-knowledge and ensemble learning.}
  \label{model_architecture}
  \end{center}
  \vspace{-1.5em}
\end{figure*}

\subsection{Parallel Context Exchange Model}

Within these distributed workflows, computing tasks $ct$ are broken down into subtasks $\tau$. These subtasks are then distributed across multiple workers $W_{R\cup C}$, with the intermediate results of subtask $\tau_n$ potentially serving as the input to the downstream subtasks $\mathbb{D}$. Hence, there arises a necessity to exchange context among the distributed computational nodes. In this paper, the geographic position of each node ($\cdot$) is mathematically characterized using latitude $\phi_{(\cdot)}$ and longitude $\lambda_{(\cdot)}$ coordinates. The spatial distance between nodes $a$ and $b$ can be calculated using the Haversine formula:

\small{
\begin{equation}
\begin{aligned}
{dist}_{a,b}(t) &= \mathbb{R} \cdot {hav}\left(\frac{\Delta\phi_{(a,b,t)}}{2}\right) \\
&\quad + \cos(\phi_a) \cdot \cos(\phi_b) \cdot {hav}\left(\frac{\Delta\lambda_{(a,b,t)}}{2}\right),
\end{aligned}
\end{equation}
}

\noindent where $\mathbb{R}$ is the Earth's radius, and ${hav}(\theta)$ is the Haversine function defined as $\sin^2(\theta/2)$. $\Delta\phi_{(a,b,t)}$ and $\Delta\lambda_{(a,b,t)}$ denote the changes in latitude and longitude between $a$ and $b$ at time $t$, respectively.

As ${dist}_{(a,b)}$ increases, it becomes more challenging to maintain reliable data transmission. According to Shannon's theorem, the transmission rate can be computed based on the Signal-to-Noise Ratio (SNR):

\small{
\begin{equation}
{tr}_{a\leftrightarrow b}(t) = \mathbb{B} \cdot \log_2\left(1 + \frac{\min\{\mathbb{P}_a, \mathbb{P}_b\} \cdot \sigma \cdot {dist}_{a,b}(t)}{\mathbb{N}_0}\right),
\end{equation}
}

\noindent where $\mathbb{B}$ is the channel bandwidth, $\mathbb{P}_{(\cdot)}$ denotes the transmission power, factor $\sigma$ is the path loss exponent, and $\mathbb{N}_0$ is the amplitude of the Gaussian background noise.

Coverage range may vary according to BSs, represented as $g_i$ with $i$ denoting a specific BS. Then, the collection of nodes capable of establishing communication with $r_i$ is denoted as $\mathbb{S}=\{x|{dist}_{(x,r_i)} \leq g_i\}$. For unreachable $r_i \notin \mathbb{S}$, mesh networking $\mathbb{M}=\{a\leftrightarrow n, \ldots, m\leftrightarrow b\}$ provides an alternative means of connectivity through relaying. As this multi-hop relay solution for two distant nodes incurs higher latency, it's typically utilized as a backup degradation in case of direct link failures. The time consumption to transmit $\tau_n$'s context from node $a$ to $b$ through $\mathbb{M}$ can then be calculated as

\small{
\begin{equation}
T_{a\leftrightarrow b}^{{mesh}}(\tau_n) = \sum_{\otimes \in \mathbb{M}} \frac{{size}(\tau_n)}{{tr}_{(a\leftrightarrow \otimes)}}.
\end{equation}
}

Data transfer between RMUs and the cloud leverages wired backhaul link on default, which provides dedicated bandwidth for long-distance data exchange. In this case, the time taken for context exchange transactions primarily depends on the Round-trip Time (RTT), which can be approximately modeled as

\small{
\begin{equation}
T_{a\leftrightarrow b}^{{backhaul}}(\tau_n) \approx {RTT} = 2 \cdot \frac{{dist}_{a,b}}{v_{{tran}} \cdot \eta},
\end{equation}
}

\noindent where $v_{{tran}}$ is the velocity of electromagnetic signal propagation, and $\eta$ represents the reduction factor due to signal attenuation within the transmission medium.

For each subtask $\tau_n$, its context exchange can occur either via multi-hop mesh network or single-hop backhaul links. Since the node's connectivity may vary, taking the minimum of these two provides an approximate optimization that always selects the low latency path. With the assumptions that $\tau_n$ only involves bidirectional context exchange at both its initiation and completion, the total end-to-end delay for a distributed task ${ct}_i$ comprising $N({ct}_i)$ subtasks can therefore be modeled as

\small{
\begin{equation}
T_{{TRANS}}({ct}_i) = \sum_{n=1}^{N({ct}_i)} \min\{T_{a\leftrightarrow b}^{{mesh}}(\tau_n), T_{a\leftrightarrow b}^{{backhaul}}(\tau_n)\}.
\end{equation}
}

\subsection{Problem Formulation}

Turnout malfunctions can result in catastrophic consequences if not addressed promptly. The real-time detection of these malfunctions is crucial for early-warning of track anomalies, providing more reaction time to train operators and maintenance groups. Our objective is to obtain an optimal policy for task partitioning and offloading that jointly optimizes execution and data transfer to meet real-time constraints. Let $\mu_E$ and $\mu_T$ be the weighting coefficients of execution time and transmission delay, where $\mu_E + \mu_T = 1$ and $\mu_E, \mu_T \in [0,1]$. The multi-objective optimization problem can be formulated as

\small{
\begin{equation}
\label{eq_opt}
\begin{aligned}
{min } & \sum_{t=1}^{T} \sum_{{ct}_i \in {CT}(t)} \mu_E \cdot T_{{EXEC}}({ct}_i) + \mu_T \cdot T_{{TRANS}}({ct}_i) \\
{s.t. } \\
C_1 & : T_{{EXEC}} + T_{{TRANS}} \leq {TIMEOUT}, \\
C_2 & : {sys\_res}(r_j) \geq \Psi, \forall r_j \in R, \\
C_3 & : {dist}_{(x,r_i)} \leq g_i, \forall r_j \in R, \\
C_4 & : T_{{start}}(\tau_n) \geq T_{{end}}(\mathbb{D}), 1 \leq n \leq N({ct}_i).
\end{aligned}
\end{equation}
}

Where $T$ represents the total run time of the system. $T_{{start}}(\tau_n)$ and $T_{{end}}(\tau_n)$ denote the start and completion time of subtask $\tau_n$, respectively. Constraint $C_1$ specifies the timeout threshold for individual tasks. Resource constraint $C_2$ guarantees that tasks can only be assigned to workers that have sufficient system resources. Distance constraint $C_3$ ensures that each BS can only communicate with devices within its coverage range. Task dependency constraint $C_4$ specifies the precedence relationships between subtasks.

\section{Parallel-Optimized Turnout Fault Diagnosis Scheme}
\label{fault_diag}

Inspired by \cite{9579018}, this paper incorporates multiple sub-models through an ensemble approach to enhance the parallelism of fault diagnosis process. As illustrated in Figure \ref{model_architecture}, the proposed model has a hierarchical modular structure comprising three main components: \textbf{a)} Segmentation module that partitions turnout operation current sequences into stages. \textbf{b)} Three parallelized fault classification models, each tailored to a particular modeling strategy. \textbf{c)} Late-fusion module to combine previous outputs and form the final result. 

\subsection{Exploiting Phase Segmentation as Prior Knowledge}

A complete turnout transition cycle comprises three distinct stages: starting, transition, and indication. Utilizing the results of stage segmentation as prior knowledge allows downstream models to conduct sequential feature extraction with enhanced effectiveness. Consequently, there exist two segmentation points $P_1$ and $P_2$, which divide the current sequence $X=\{x_1,x_2,\ldots,x_n\}$ into $X_{Stage1}$, $X_{Stage2}$, and $X_{Stage3}$. Ou et al. \cite{doi:10.1177/0361198119837222} leverages second-order difference to identify $P_1$ and $P_2$ in $X$, as it's particularly sensitive to these inflection points. However, the intense current fluctuations in faulty samples can easily exceed the generalization capabilities of traditional numerical-based algorithms in handling variations, thereby impacting segmentation accuracy. To address this, we employ a GRU (Gated Recurrent Unit) network to analyze three-phase current sequences (A, B, C channels) and assign confidence scores reflecting the likelihood of each point denoting a segmentation boundary. The final confidence score for each potential segmentation point can be computed as

\small{
\begin{equation}
\begin{aligned}
{Score}(i) = & \frac{|x_i - {mean}(X_{{Stage2}})|}{|x_i - {mean}(X_{{Stage3}})|} * d_i W_i \\ & + \gamma [{Score}(i+1) + {Score}(i-1)] \\
& + \gamma^2 [{Score}(i+2) + {Score}(i-2)] + \cdots,
\end{aligned} 
\end{equation}
}

\noindent where ${mean}(\cdot)$ denotes the stage average, $d_i$ represents the height of the potential peak, and $W_i$ is the GRU confidence score output. The discount factor $\gamma$ assigns lower influence to distant peaks, thus emphasizing local relationships while leveraging global dependencies.

The proposed segmentation scheme operates independently on the three-phase current channels. Consequently, selecting the highest score yields three sets of candidate segmentation points $\{<P_1^A, P_2^A>, <P_1^B, P_2^B>, <P_1^C, P_2^C>\}$. Ideally, segmentation points denoting the same boundary (e.g., $\{P_1^A, P_1^B, P_1^C\}$) should exhibit close agreement if identified correctly. To reconcile such multi-channel results, an outlier detection algorithm \cite{10.1145/3444690} is introduced to discard anomalous points. Ultimately, the remaining healthy points are averaged to obtain the final output.

\subsection{Three-Stage Feature Extraction and Fusion}

In pursuit of a fault classification paradigm exhibiting robustness, parallelizability and interpretability, we adopt an ensemble approach that integrates predictions from distinctive sub-models. Specifically, three sub-classifiers based on \textbf{a)} time-domain feature engineering, \textbf{b)} morphological similarity, and \textbf{c)} deep feature extraction, are developed to address fault classification from different perspectives.

\textbf{a) Multi-layer Perceptron (MLP):} As a Neural Network (NN) model, MLP demonstrates remarkable fitting and generalization capabilities, making it well-suited for classification problems \cite{10.1093/tse/tdac036}. Time-domain features (Table \ref{feature_engineering}) are carefully selected to construct feature set $F_{{Stage}\_x}$ for each segmented stage sequence. These sets are then normalized and combined to form the comprehensive stage-wise features set $\langle F_{{Stage}\_1}, F_{{Stage}\_2}, F_{{Stage}\_3} \rangle$, which serves as the input for the model. When fewer than two segmented points are recognized, all feature values corresponding to the missing stages will be set to -1. Additionally, when encountering divide-by-zero during feature extraction, the output will be set to 0 to prevent triggering an exception.

\begin{table}[h]
\renewcommand{\arraystretch}{1.7}
\caption{Time-domain features extracted for each stage segment (partial).}
\label{feature_engineering}
\centering
\begin{tabular}{>{\raggedright}m{0.22\columnwidth} >{\centering}m{0.38\columnwidth} >{\raggedright}m{0.25\columnwidth}}
\hline
\textbf{Feature Type} & \textbf{Calculation Formula} & \textbf{Description} \\
\hline
Peak-to-Peak & $X_{{max}} - X_{{min}}$ & Amplitude range \\

Std & $\sqrt{\left(\sum_{i=1}^n \left(\frac{(x_i - \bar{X})^2}{n}\right)\right)}$ & Signal stability \\

Kurtosis & $\frac{\sum_{i=1}^n \left(\frac{(x_i - |\bar{X}|)}{{std}}\right)^4}{n}$ & Distribution shape\\

Clearance Factor & $\frac{X_{{max}}}{\left(\sum_{i=1}^n \sqrt{|x_i|})/n\right)^2}$ & Separation extent\\
\hline
\end{tabular}
\end{table}

\textbf{b) Denoising Auto Encoder (DAE):} DAE performs non-linear dimensionality reduction while extracting higher-level descriptors of waveform shape, showcasing wide applications in unsupervised time-series anomaly detection \cite{WOS:000966540400001}. The morphological characteristics of current waveforms, such as subtle variations and local extrema distributions, are challenging to capture numerically. However, such features are proved useful for determining fault types. The DAE operates on a four-dimensional input: three channels allocated for phase current sequences, complemented by a binary segmentation mask channel that uses boolean values (0 and 1) to identify segmentation points. Subsequently, Mean Absolute Error (MAE) is employed to form the total reconstruction error set $L_{ae} = \{L_{ae}^{Normal}, L_{ae}^{H1}, \ldots, L_{ae}^{F5}\}$. Each element in $L_{ae}^{type}$ consists of losses from the three stages, represented as $L_{ae}^{type} = \langle l_{Stage1}, l_{Stage2}, l_{Stage3} \rangle$. Larger loss indicates more significant morphological differences, suggesting lower confidence that the sample belongs to that category. Contributions from each stage are aggregated using weight assignments to compute the anomaly score $\tilde{S}_{ae}$. Subsequently, a numerical inversion and scaling of $\tilde{S}_{ae}$ yields the fault type classification confidence $c_k$ ($k \in \{ {Normal}, H1, \dots, F5 \}$), calculated as

\small{
\begin{equation}
c_k = {Softmax} \left( \frac{e^{-m\tilde{S}_{ae}^k}}{\left(1 + e^{-m\tilde{S}_{ae}^k}\right)^2} \right),
\end{equation}
}

where $m$ is the scaling coefficient and $Softmax()$ normalizes the output to ensure a valid probability distribution across fault types.

\textbf{c) Temporal Convolutional Network (TCN):} As a one-dimensional Fully Convolutional Network (FCN) designed specifically for sequential data, TCN \cite{WOS:000982475600005} is regarded as the successor to Recurrent Neural Networks (RNNs). The model's architecture features a four-channel input, consistent with the DAE sub-classifier, and incorporates a linear layer for direct classification result output.

Fuzzy Logic (FL) is a computational paradigm where a value can belong to multiple fuzzy sets, each associated with a membership degree ~\cite{WOS:000848264000029}. In this paper, we leverages FL for combining results from multiple sub-classifiers at the decision level. Specifically, we perform fuzzy modeling of outputs from individual classifiers to account for ambiguity inherently associated with classification problems. Membership functions gauge membership to fuzzy set $\{Negative,Positive\}$ defined over the domain of all fault categories. For a given classifier, Algorithm  \ref{fuzzy_fusion} outlines the process of determining membership functions for each domain with statistical experimental method. The three-dimensional output array $W$ has fuzzy domains on its first dimension and fuzzy sets on the second, with its elements on the third dimension mapping to the corresponding membership function $\mu(x)$.

\begin{algorithm}
\small 
\caption{Determine Membership Functions for Each Classifier}
\label{fuzzy_fusion}
\textbf{Input:} Classifier $C$, Dataset $D$, Sample types $F$, Stride $t$, Number of folds $k$ \\
\textbf{Output:} Membership Functions $M$

type({$q$}) //return the fault type $q$ belongs to\\

Split $D$ into $k$ sets $S=\{S_1,S_2,\ldots,S_k\}$ \\
Divide possibility range 0 to 1 with stride $t$ into $R$ \\
Initialize matrix $M$ of shape $(size ~ of ~ F,2,size ~ of ~ R)$ \\

\For{\textbf{each} $S_i$ \textbf{in} $S$}{
    Select $S_i$ as the validation set \\
    Train $C$ on the remaining $k-1$ sets \\
    \For{\textbf{each} sample $s$ \textbf{in} $S_i$}{
        Obtain classifier output $Q$ consisting of confidences to each category \\
        \For{\textbf{each} confidence score $q$ \textbf{in} $Q$}{
            $h =$ index of the range $q$ belongs to in $R$ \\
            $M[\text{index of type}(q) ~ \text{in} ~ F][\text{type}(q)==\text{type}(s)][h]++$ \\
        }
    }
}

\For{$i=0$; $i<$ size of ~$F$; $i++$}{
    \For{$j=0$; $j<2$; $j++$}{
            Perform standardization to each element using $\frac{M[i][j][k]}{sum(M[i][j])}$ \\

    }
}

\textbf{return} $M$

\end{algorithm}

\begin{figure*}[b]
  \begin{center}
  \includegraphics[width=0.7\linewidth]{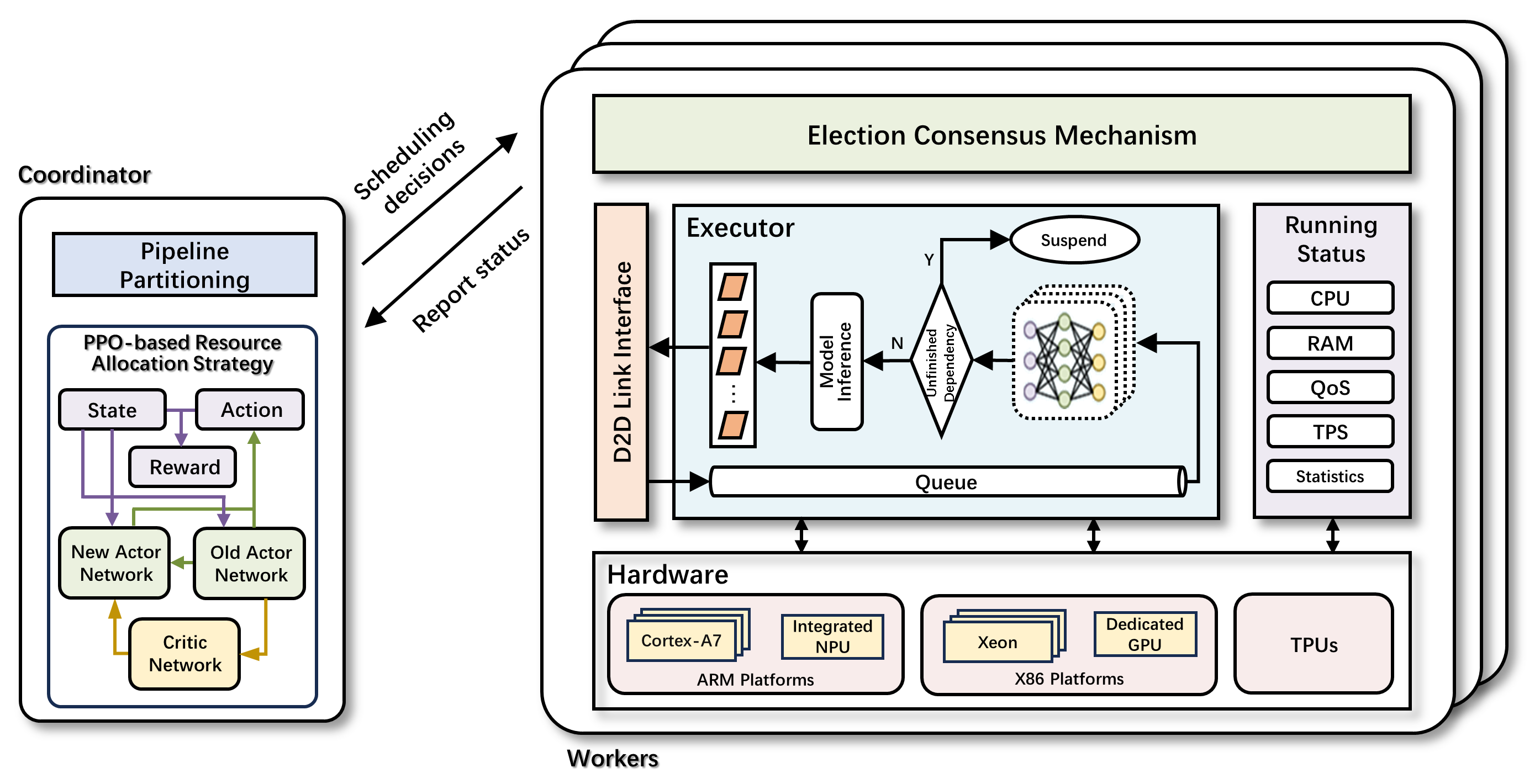}
  \caption{The framework of CEC-PA.}
  \label{CEC-PA}
  \end{center}
\end{figure*}

\begin{figure*}[b]
  \begin{center}
  \includegraphics[width=0.6\linewidth]{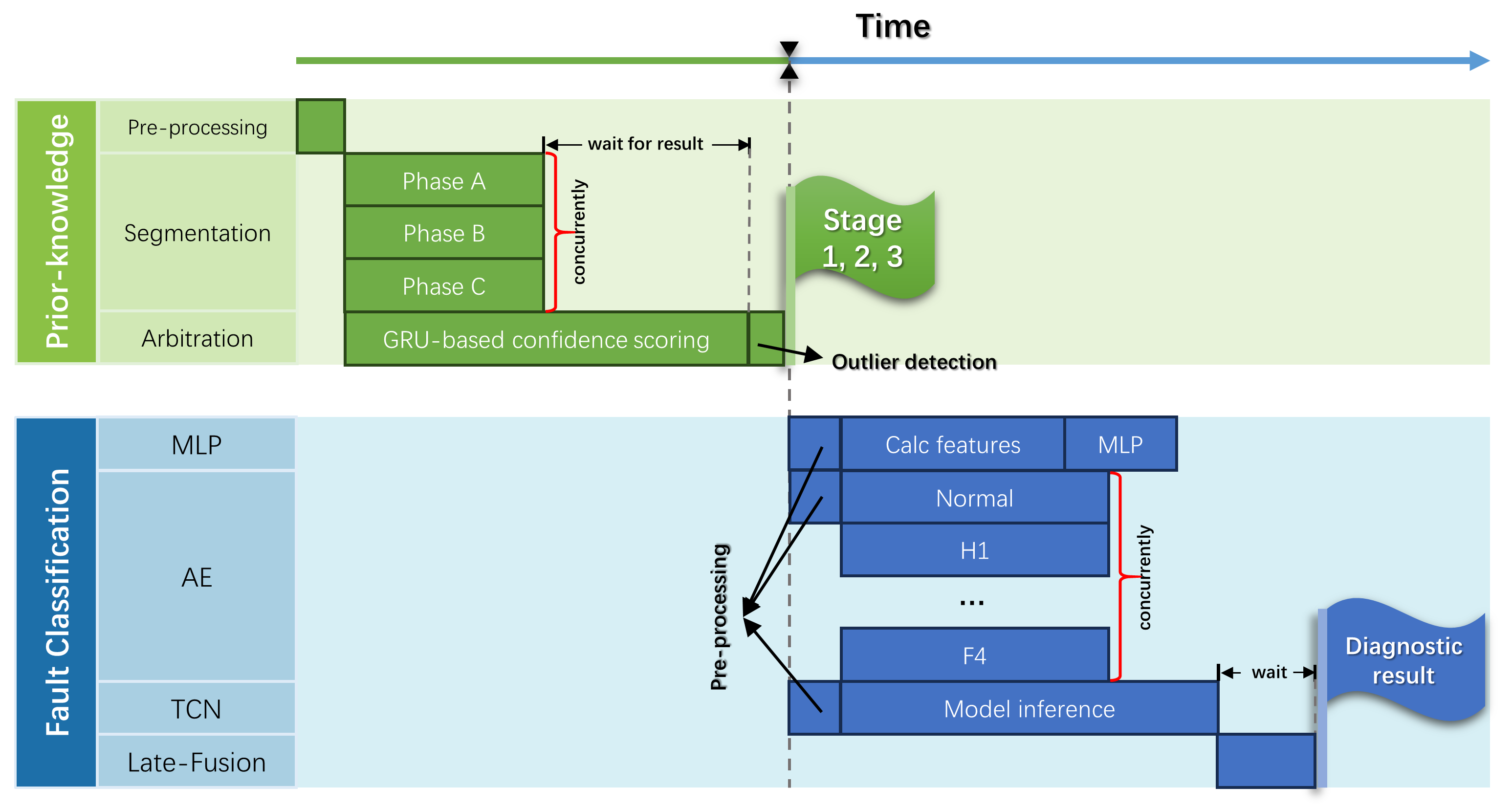}
  \caption{Gantt chart representation of fault diagnosis model inference workflow template decomposed at the atomic model component level.}
  \label{Gantt}
  \end{center}
  \vspace{-1em}
\end{figure*}

Let $x_i^t$ denote the confidence level assigned by classifier $i$ to the input sample being of fault type $t$, and $\mu_{i,j}$ represent the membership function of classifier $i$ on domain $j$. The classification confidence for each fault category is individually mapped through the corresponding membership functions, yielding membership degrees $y_{i,t}=\mu_{i,t}(x_i^t)$. Subsequently, a comprehensive membership degree $\hat{y}_{i,t}$ is computed by subtracting the membership degree associated with "Negative" from that associated with "Positive":

\small{
\begin{equation}
\hat{y}_{i,t} = y_{i,t}[{Positive}] - y_{i,t}[{Negative}].
\end{equation}
}

To obtain the final classification result $Y$, the Softmax function is applied to fuse the membership vectors from all available classifiers:

\small{
\begin{equation}
Y = {Softmax}(\hat{y}_{MLP} + \hat{y}_{DAE} + \hat{y}_{TCN}).
\end{equation}
}

\section{CEC-PA: A Cloud-Edge Collaborative Pipeline Parallelism Framework for Distributed Fault Diagnosis}
\label{section_cec}

Monolithic implementations of hybrid fault diagnosis models, where prior knowledge extraction, sub-classifiers, and late-fusion module are executed on a single centralized node, typically exhibit inefficient resource allocation and compromised system responsiveness. To address these limitations, the Cloud-Edge collaborative parallelism-aware scheduling framework, namely CEC-PA, is proposed to intelligently schedule tasks across worker nodes in a parallelized manner. Specifically, CEC-PA operates in conjunction with the previous hierarchical diagnosis model, which is now partitioned at a fine-grained level to fully exploit the distributed computational capabilities across cloud and edge, as depicted in Figure \ref{CEC-PA}.

In this section, a partitioning strategy is first presented to divide the overall fault diagnosis model into pipelines. Then, the pipeline offloading problem is formalized as a Markov Decision Process (MDP) and a DRL-based computation offloading policy is introduced to output optimal pipeline-worker mappings in response to the dynamic environment.

\subsection{Parallel Task Partitioning Across Pipelines}
\label{section_6.1}
As discussed in section \ref{section_4.2}, parallel tasks may have dependencies on the outputs of prior tasks. Allowing minimal units to be directly scheduled may result in task accumulation and blocking across pipelines. Therefore, assigning coupled tasks to the same pipeline is the key to reduce overheads. We choose to partition at the pipeline-level, rather than neuron-level by the fact that model inference involves both computationally intensive operations and substantial memory access patterns. In contrast, neuron-level parallelism approaches \cite{10177476} rely heavily on low latency and high bandwidth network environments. The black-box nature at the model component level provides good isolation by exposing only the inputs and outputs. This characteristic aligns seamlessly with our distributed pipeline parallelism approach, where our aim is to minimize context exchange and data throughput for computation tasks scheduled across the network.

As depicted in Figure~\ref{Gantt}, the proposed fault diagnosis workflow comprises both sequentially dependent and parallelizable components. The classification exhibits dependency on prior-knowledge segmentation results and requires strict serialization. The arbitration of segmentation points and the ensemble of classifiers need to wait for their predecessor tasks to finalize. However, segmentation of individual current phases and the inference of sub-classifiers (i.e., MLP, AE, TCN) are independent of each other and can proceed fully in parallel. Let the Directed Acyclic Graph (DAG) be represented as $G = (V, E)$, where $V$ is the set of subtasks and $E$ is the set of dependency edges. Consolidating tasks with minimal or short-term dependencies is paramount for preserving intrinsic parallelism within the system. Transitive dependency chain between non-adjacent tasks $v_i$ and $v_j$ can be identified using path connectivity, where $\exists$ path $\in E$: $v_i \rightarrow \ldots \rightarrow v_j$ implies $v_j$ transitively depends on $v_i$.

Additionally, tasks with similar resource demand patterns should be co-located to optimize resource utilization. The resource profile for each task is represented as a multivariate vector $\Psi$. A greedy partitioning strategy is then employed with the granularity parameter $G$. The algorithm initializes a given number of $G$ pipelines and iterates through tasks in topological order. Eventually, each $v_i$ will be assigned to the pipeline $P_j$ that maximizes the affinity function, which can be expressed as the product of resource pattern similarity and task dependency score:

\small{
\begin{equation}
 G(v_i, P_j) = \sum_{v_k \in P_j} \frac{\Psi_i \cdot \Psi_k}{||\Psi_i|| \cdot ||\Psi_k||} \cdot \sum_{P^* \in \{P - P_j\}} \ln[v_* \rightarrow v_i],
\end{equation}
}

\noindent where resource pattern similarity is measured by the cosine similarity between resource profile vectors $\Psi_i$ and $\Psi_k$. Dependency score is calculated based on the path length $v_* \rightarrow v_i$ between $v_i$ and tasks assigned to pipelines other than $P_j$.

\subsection{Formulation of Markov Decision Process}

Once the partitioning of model components has been determined, the resultant pipelines must then be properly offloaded onto available worker nodes (i.e., Cloud center $W_C$ or RMU edges $\{W_{r_1}, W_{r_2}, ..., W_{r_J}\}$ to minimize the total time consumption. Traditional scheduling algorithms such as Round Robin and First Come First Serve (FCFS) fall short in this context due to the dynamic nature of the network and high-dimensional state space arising from the cloud-edge environment. To tackle these challenges, DRL-based task offloading approaches \cite{9590352} \cite{9284332} are proposed. By continuously interacting with the environment and maximizing cumulative rewards, DRL agents can adjust policy $\Pi$ to enable real-time and adaptive computation offloading. The decision-making process of agents is formulated as a Markov Decision Process (MDP), which comprises:
\textbf{a) State Space:} The state space $\mathcal{S}$ encapsulates key observations about the environment to form the foundation for agents’ decision-making. Its design jointly considers properties of the distributed pipelines and real-time status of the worker nodes to comprehensively reflect the overall environment. Based on its pending subtasks, property signature $\mathcal{S}_{{par}}$ of pipeline $P_j$ includes: \textbf{1) Priority Compensation Factor:} Scaling factor that exponentially escalates $P_j$'s priority based on its waiting time, calculated as \small{$e_{}^{t_{{now}} - t_{P_j}^{{birth}}}$}. This fosters responsiveness for pipelines that have experienced prolonged queuing delays. \textbf{2) Minimum Environment Requirement:} The fundamental system resources required for execution, denoted as $\Psi_{P_j}$. \textbf{3) Dependency Encoding:} Obtained by transforming dependencies of its predecessors into a high-dimensional vector using a Graph Neural Network (GNN) encoder. For each worker node $W_i$, we capture its status $\mathcal{S}_{{node}}$ as: \textbf{1) D2D Connection Type $\gamma$:} A state quantity indicating the communication capability as either wired ($\gamma=0$) or wireless ($\gamma>0$). In cases of wireless communication, $\gamma$ additionally indicates the number of hops \cite{WOS:000510677500008} within the connection. \textbf{2) Workload:} The level of computational burden quantified as $\frac{{UP\_TIME}}{T_{{COMP}}}$, where $T_{{COMP}}$ represents the total time slices in non-idle state. This metric is critical for load balancing and resource allocation. \textbf{3) Hardware Metadata:} Information on whether the node is equipped with Application-Specific Integrated Circuits (ASICs) for hardware acceleration, such as GPU, NPU, and TPU. \textbf{4) Link Quality:} This metric is defined to be proportional to the average throughput $tr_{*\leftrightarrow W_i}$ and inversely proportional to the packet loss rate $\sigma$, which can be denoted as $\frac{\ln(1+tr_{*\leftrightarrow W_i})}{\sigma^2}$. Hence, the two subsets $\{\mathcal{S}_{{par}}, \mathcal{S}_{{node}}\}$ are bundled together to form a holistic representation of pipeline $P_j$ and worker node $W_i$.

\textbf{b) Action Set:} Given a state observation encoding $\mathcal{S}$, the policy network outputs $\Pi_{\theta}(\mathcal{S})$ for manipulating whether to offload $P_j$ onto $W_i$. The action set $\mathcal{A}$ encompasses all potential actions that the agent scheduler can take, defined as $\mathcal{A}=[\delta, \eta]$. Here, $\mathcal{A}$ is expressed as a discrete action space, so that the agent can only select one action at a time, denoted as $\delta, \eta \in \{0,1\}$ with $\delta + \eta = 1$. $\delta$ and $\eta$ are Boolean variables that represent the idle and offload actions, respectively. The selection result $\Pi_{\theta}(\mathcal{S})$ should be either $\delta = 0$ and $\eta = 1$, indicating that the agent offloads $P_j$ onto $W_i$, or $\delta = 1$ and $\eta = 0$, indicating that the agent idles and skips $P_j$.

\textbf{c) Reward Function:} According to Equation (\ref{eq_opt}), our objective is to jointly minimize the execution time and transmission delay. Additionally, to avoid the agent getting stuck in local optima, a success rate term $\sigma$ is introduced for penalizing constraint violations. The overall reward increases as $\sigma$ decreases, thereby motivating the agent to align its actions with real-world scenarios. The final reward function $\mathcal{R}$ is constructed in a way that it is intended to be minimized, which is presented as

\small{
\begin{equation}
R = -\frac{\mu_E \cdot T_{{EXEC}} + \mu_T \cdot T_{{TRANS}}}{\ln(\sigma)}.
\end{equation}
}

\subsection{PPO Empowered Computation Offloading for Pipelines}

Given the dynamic and complex nature of cloud-edge systems, solving the resulting Markov decision process (MDP) directly is computationally intractable as it belongs to NP-hard complexity. Traditional Reinforcement Learning (RL) algorithms such as Q-learning and Sarsa hinge on tabular representations between states and actions, which are proven impractical when confronting expansive state spaces. Meanwhile, algorithms such as Deep Deterministic Policy Gradient (DDPG) encounter limitations in handling discrete action sets. More advanced algorithms like Twin Delayed Deep Deterministic Policy Gradient (TD3) may impact the responsiveness of intelligent offloading decisions due to their resource-intensive network structures \cite{WOS:001059956100001}. In this context, Proximal Policy Optimization (PPO) emerges as a discerning choice. PPO's efficacy in handling high-dimensional state spaces and discrete action sets, coupled with its runtime adaptability, distinguishes it among its counterparts.

PPO employs an Actor-Critic architecture, where the actor network $\theta_A$ is responsible for interacting with the environment and the critic network $\theta_C$ evaluates the actions taken by the actor. At each timestep, the agent observes state $S$ and selects an action from $\theta_A$’s policy $\Pi_{\theta_A}(\cdot|\mathcal{S})$. The environment transitions to $\mathcal{S}'$, and a reward $\mathcal{R}$ is received. This experience $\langle \mathcal{S}, \mathcal{A}, \mathcal{R}, \mathcal{S}' \rangle$ is stored in the replay buffer $\textit{buff}$. Periodically, the networks are trained with minibatches sampled from the buffer. The value loss enforces $\theta_C$ to match the observed returns, while the policy loss employs a clipping mechanism to maintain stability. The surrogate objective which guides these policy updates can be defined as:

\begin{equation}
\label{EQ19}
L^{\theta'}(\theta) = E\left[\min\left(r^{\theta'}(\theta) \cdot \hat{\lambda}, clip \left(r^{\theta'}(\theta), 1 - \epsilon, 1 + \epsilon\right) \cdot \hat{\lambda}\right)\right],
\end{equation}

\noindent where $r^{\theta'}(\theta)$ is the possibility ratio of new and old policies, calculated as $\frac{\Pi_{\theta}(\mathcal{A}|\mathcal{S})}{\Pi_{\theta'}(\mathcal{A}|\mathcal{S})}$. Additionally, $\hat{\lambda}$ is the advantage function, and $\epsilon$ is a hyper-parameter controlling the degree of policy change. The term $clip(\cdot)$ limits the policy update to be within a certain range, preventing excessively large updates.

During each iteration, the policy network parameter is updated using gradient ascent to maximize $L^{\theta'}(\theta)$. Let ${KL}$ represent the Kullback-Leibler divergence, which serves as a regularization term to ensure the new policy $\theta'$ does not deviate too far from the old policy $\theta$ during updates. The hyper-parameter $\beta$ adjusts the impact of the KL divergence. Then, the objective function can be defined as

\small{
\begin{equation}
\label{EQ20}
J_{PPO'}(\theta) = \arg\max_{\theta} \left( L^{\theta'}(\theta) - \beta \cdot {KL}(\theta, \theta') \right).
\end{equation}
}

\begin{algorithm}[h]
\small 
\caption{PPO-empowered Computation Offloading Decision Process}
\label{ppo_computation_offloading}
\textbf{Input:} Actor network $\theta_A$, Critic network $\theta_C$, Worker nodes $W$, Pipeline $P$ \\
\textbf{Output:} Offloading decisions $\mathcal{A}_{\tau_j \rightarrow W_i} (\tau_j \in P,  W_i \in W$) \\
Initialize network parameters $\theta_A$ and $\theta_C$ \\
Initialize replay buffer \textit{buff}

\For{each subtask $\tau_j$ in $P$}{
    /* Skip to the next subtask if $\tau_j$ has been completed */ \\
    \If{$\tau_j.\Phi$ is completed}{
        \textbf{continue}
    }
    \For{each worker $W_i$ in $W$}{
        Observe $\mathcal{S}_{par}$ and $\mathcal{S}_{node}$ from the environment to construct current state $\mathcal{S}$ \\
        Select $\mathcal{A}_{\tau_j \rightarrow W_i}$ based on $\Pi_{\theta_A}(\cdot|\mathcal{S})$ \\
        Take action $\mathcal{A}_{\tau_j \rightarrow W_i}$ on $W_i$, and capture exception if it violates constraints \\
        Obtain the next state $\mathcal{S}'$ and calculate the reward $\mathcal{R}$ \\
        Store the new experience $\langle \mathcal{S}, \mathcal{A}, \mathcal{R}, \mathcal{S}' \rangle$ into \textit{buff}

        Randomly select $k$ on-policy mini-batches from \textit{buff} \\
        \For{$i=0$ to $k$}{
            Calculate the surrogate objective based on $\theta$ and $\theta'$ using Equation (\ref{EQ19}) \\
            Update $\theta_A$ and $\theta_C$ with gradient ascent using Equation (\ref{EQ20}) \\
        }
        Save the updated network parameters: $\theta' \leftarrow \theta$
    }
}
\end{algorithm}

Algorithm \ref{ppo_computation_offloading} elaborates the iterative decision-making process of agents. Lines 1-2 initialize the variables before execution. Line 4 iterates through all subtasks in the pipeline for scheduling decisions. By skipping subtasks that are already completed (line 9), the algorithm optimizes performance by mitigating redundant computations. Lines 12-16 involve action selection and execution. Notably, there is no explicit check and early-exit declaration after line 14, which allows a subtask to be concurrently scheduled across multiple nodes. This strategically designed behavior reduces queuing delays for faster response, meanwhile serving as an extra layer of safety against network congestion. Lines 18-24 sample experiences from the replay buffer to update the policy model weights, which enables the policy to continuously evolve and adapt to the dynamic environment.

\subsection{Downtime Tolerance Mechanism for the Coordinator Node}

In our proposed turnout fault early-warning system, cloud center is selected on default as the centralized coordinator which performs the CEC-PA scheduling scheme. This is mainly because achieving consistency in distributed systems \cite{WOS:000833520600006} has always been a complex research challenge. Having multiple coordinators introduces the risk of race conditions and inconsistent scheduling behaviors. Nevertheless, cloud center typically possesses greater computational resources compared to edge nodes, rendering it better suited for this task. However, relying on a centralized coordinator poses a Single Point of Failure (SPOF) risk. If the cloud center experiences an unexpected outage, the entire system would become unavailable. Given the mission-critical nature of the railway system, any downtime can potentially lead to disastrous consequences.

\begin{figure}[h]
\centering
\begin{minipage}[t]{0.5\textwidth}
   \centering
\subfloat[Role transition graph with corresponding events.]{
		\includegraphics[width=0.8\columnwidth]{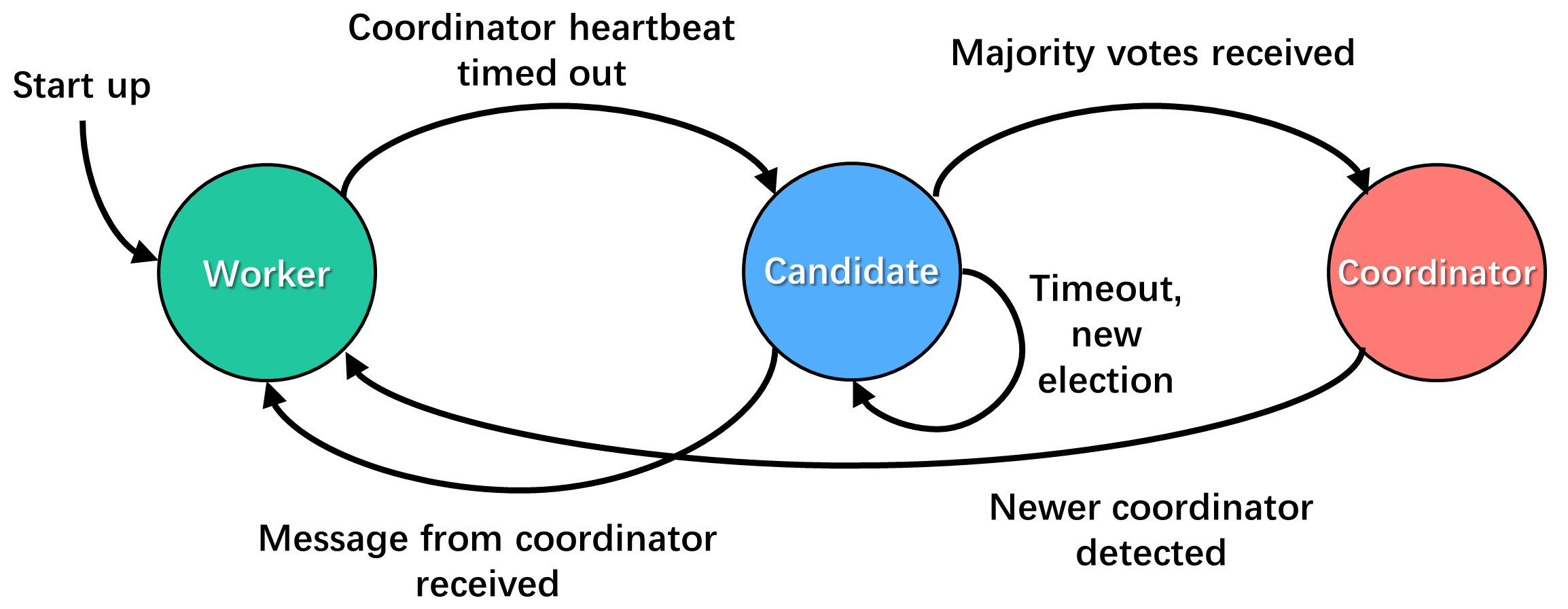}
    }\\
\subfloat[UML sequence diagram of coordinator election process.]{
		 \includegraphics[width=0.85\columnwidth]{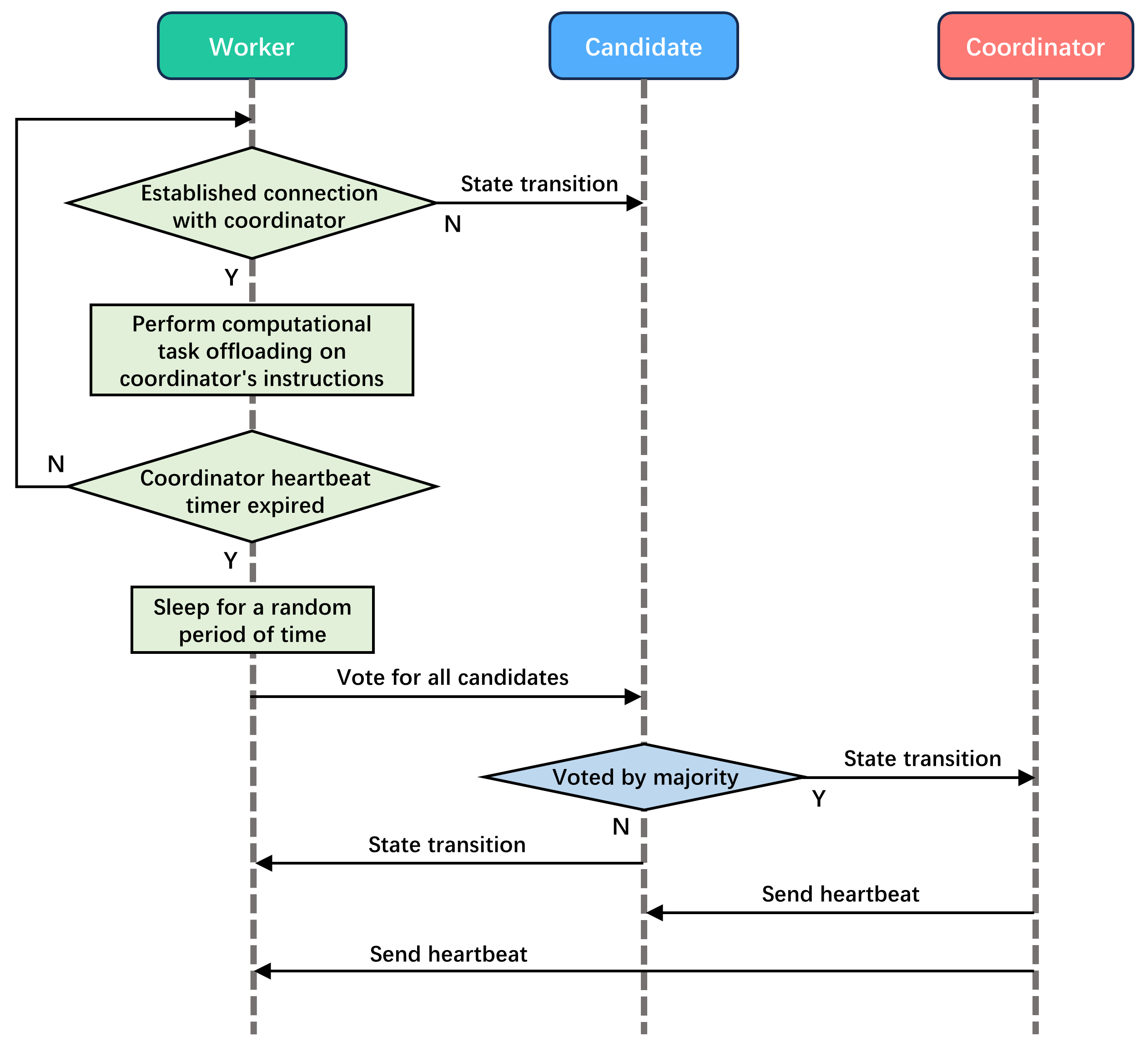}
    }
\end{minipage}
  \caption{Implementation of the proposed downtime tolerance mechanism.}
  \label{fault_tolerance}
\end{figure}

Inspired by distributed consensus protocols such as Paxos, Raft \cite{10.5555/2643634.2643666} and Gossip, a downtime tolerance mechanism (Figure \ref{fault_tolerance}) is proposed. In the event of cloud outages, a replacement coordinator is quickly elected by the consensus of the remaining nodes. Figure \ref{fault_tolerance}(a) illustrates the nodes’ transition process among the roles of coordinator, worker, and candidate. Among these roles, workers passively respond to the coordinator's instructions. The coordinator periodically sends heartbeats to its workers, signifying its online status. If a worker fails to receive a heartbeat confirmation within timeout, it then transitions to the candidate state and initiates a coordinator election. Figure \ref{fault_tolerance}(b) depicts the election process. Each candidate waits for a randomized duration before requesting votes from other nodes. This fundamentally consistency errors that could arise from a competing condition. Once a majority of votes are acquired, the candidate gets promoted to the new coordinator and broadcasts its updated identity to all nodes in the network.

\begin{figure*}[b]
  \centering
  \captionsetup[subfigure]{oneside,margin={0.7cm,0cm}}
  \subfloat[Losses of MLP.]{\includegraphics[width=0.333\linewidth]{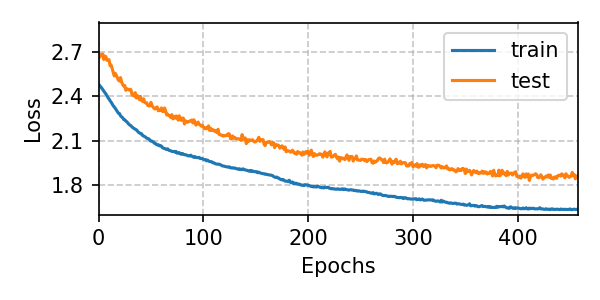}}
  \subfloat[Losses of DAE.]{\includegraphics[width=0.333\linewidth]{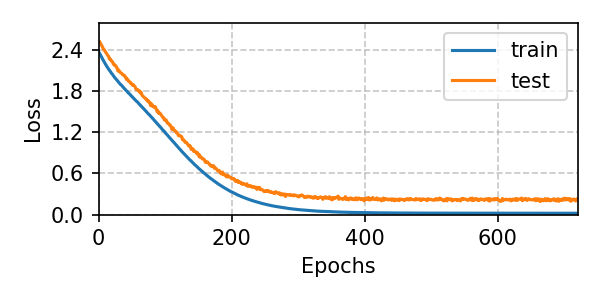}}
  \subfloat[Losses of TCN.]{\includegraphics[width=0.333\linewidth]{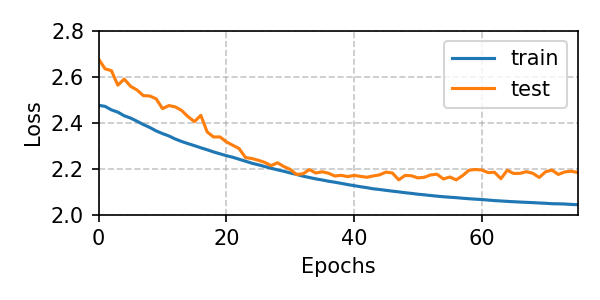}}\\
  \captionsetup[subfigure]{oneside,margin={0cm,0cm}}
  \centering
  \subfloat[Average reward per episode of PPO.]{\includegraphics[width=0.45\linewidth]{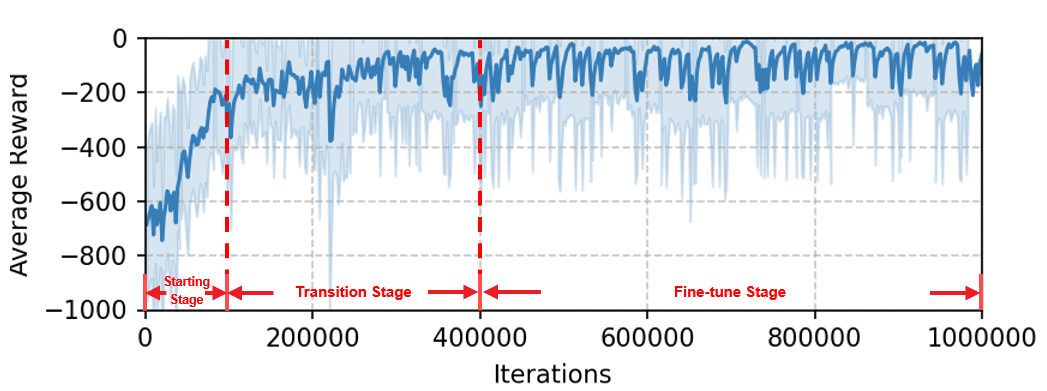}}
  \subfloat[Number of surviving episodes of PPO.]{\includegraphics[width=0.45\linewidth]{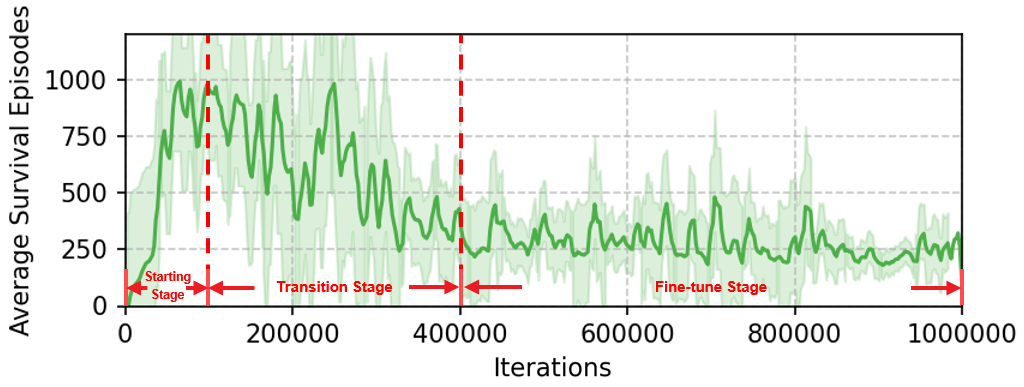}}
  \caption{Convergence metrics of integrated diagnosis and pipeline scheduling models.}
  \label{converge}
\end{figure*}

\section{Experimental Results and Analysis}
\label{experiments}
\subsection{Dataset Description and Simulation Setup}
The dataset used in this study is provided by the Nanjing Metro Bureau located in Jiangsu Province, China. The deployment of RMUs throughout the Nanjing Metro system embodies a significant technological challenge, involving intricate communication networks, power supply issues, and carefully orchestrated construction during service interruptions. This initiative has successfully transitioned from a single-line pilot program in 2021-2022 to a multi-line implementation phase in 2023. The monitored RTMs consist of Siemens S700K models and their replicated version ZD6, both operating on three-phase AC power input and sharing the same operation patterns. Spanning from November 2021 to September 2023, the dataset encompasses 10,000 samples collected from 227 turnouts deployed across 12 lines. Notably, the bureau applied data augmentation techniques \cite{10.1007/978-981-13-7542-2_15} \cite{10113353} to address the common issue of class imbalance in fault diagnosis datasets, where fault samples are typically fewer than normal samples. As a result of these efforts, the dataset provided subsequently exhibits a well-balanced distribution of labels across all categories, as illustrated in Table \ref{tab:dataset}. For each sample, the recorded information include timestamp, turnout id, three-phase current sequences with a sampling rate of 25Hz, communication quality, GPS coordinates, and rotation direction. The dataset is annotated with one normal type and 11 typical fault types, categorized into two levels based on severity: hidden dangers (H1-H6) and critical faults (F1-F5). Hidden dangers typically have no impact on normal operations but indicate a need for maintenance. However, critical faults involve rather serious issues like control circuit errors, mechanical faults, or degraded components that already lead to malfunctions. The dataset is divided into training and testing sets with an 80\% - 20\% ratio. 

\begin{table}[h]
\renewcommand{\arraystretch}{1.4}
\centering
\caption{Label Distribution of Nanjing Metro Dataset}
\centering
\begin{tabular}{cccc}
\hline
\textbf{Sample Type} & \textbf{Train Set} & \textbf{Test Set} & \textbf{Total Samples} \\
\hline
\textbf{Normal} & 760 & 190 & 950 \\
\textbf{H1}     & 736 & 184 & 920 \\
\textbf{H2}     & 680 & 170 & 850 \\
\textbf{H3}     & 668 & 167 & 835 \\
\textbf{H4}     & 640 & 160 & 800 \\
\textbf{H5}     & 624 & 156 & 780 \\
\textbf{H6}     & 632 & 158 & 790 \\
\textbf{F1}     & 656 & 164 & 820 \\
\textbf{F2}     & 620 & 155 & 775 \\
\textbf{F3}     & 656 & 164 & 820 \\
\textbf{F4}     & 660 & 165 & 825 \\
\textbf{F5}     & 668 & 167 & 835 \\
\hline
\end{tabular}
\label{tab:dataset}
\end{table}

The experiments are conducted on an Ubuntu 20.04.3 LTS server with Intel i7-12700KF CPU (3.6 GHz, 20 cores), 64 GB DRAM, and a NVIDIA RTX 4090 GPU. The runtime environment is configured with PyTorch 2.0.1 and Python 3.11.3. To simulate the Cloud-Edge network architecture, a modified version of VEC-Sim \cite{wu2024vec} is implemented with cloud-edge collaborative support and RTM fault diagnosis model embedded. The parameters in Table \ref{sim_param} are selected based on the actual conditions of the Nanjing Metro system, with a focus on reflecting the operational scenario once the RMUs are fully deployed.

\begin{table}[h]
\renewcommand{\arraystretch}{1.5}
\caption{Simulation Parameters}
\label{sim_param}
\centering
\begin{tabular}{>{\raggedright}p{0.4\columnwidth} >{\raggedright}p{0.35\columnwidth}}
\hline
\textbf{Parameter} & \textbf{Value} \\
\hline
Number of Turnouts $S$ & 1000 \\
Turnout Operation Interval & 10 -- 30 min \\
Cloud Computation Capacity & 200 GFLOPS \\
Number of RMUs $J$ & 50 \\
RMU Computation Capacity & 1 GFLOPS -- 10 GFLOPS \\
RMU Queue Size & \{4, 8, 16\} randomly \\
RMU Coverage Range $g_i$ & 150m -- 2km \\
Task Computation Distribution & 2 -- 10 GFLOPS (Zipf) \\
Task Dependency Possibility & 0.5 \\
D2D Bandwidth $\mathbb{B}$ & 300 Mbps \\
Request Timeout Limit & 10s \\
\hline
\end{tabular}
\end{table}
\vspace{-1em}

\subsection{Model Convergence Analysis}

In the proposed fault diagnosis scheme, multiple sub-classifiers (i,e., MLP, DAE and TCN) are integrated to jointly analyze the input sample. The potential failure of any sub-classifier can impact the final diagnosis result. Furthermore, the effectiveness of the CEC-PA scheduling framework relies on utilizing MDP to capture pertinent environmental dynamics. In cases where the MDP is inaccurate or deficient, PPO agents may struggle to discover effective policies, leading to non-convergence. To validate the stable and synergistic operation of both the diagnostic model and the offloading decision-making model, their convergence profiles during training are illustrated in Figure \ref{converge}.

\begin{figure*}[b]
\centering
 \centering
 \begin{minipage}[t]{0.85\textwidth}
   \centering
   \subfloat[Fault samples encoded by matched DAEs.]{\includegraphics[width=\linewidth]{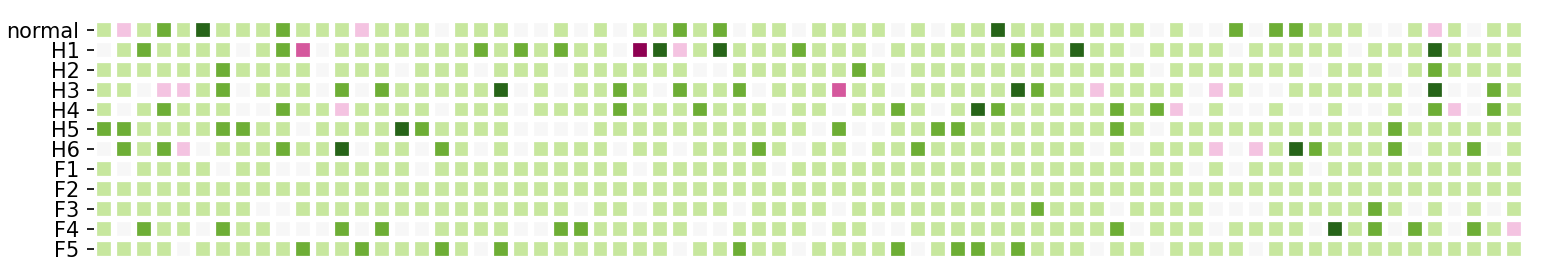}}
   \hfill
   \subfloat[Fault samples encoded by DAEs trained on normal samples.]{\includegraphics[width=\linewidth]{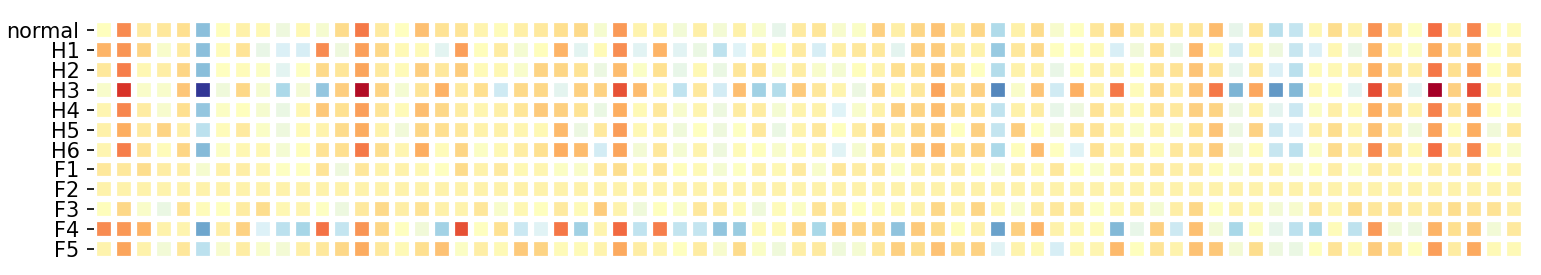}}
 \end{minipage}
 \caption{Latent vector output of DAE’s bottleneck layer.}
 \label{dae_visualization}
 \vspace{-2em}
\end{figure*}

The training hyper-parameters for all models are standardized with batch size of 128 and learning rate of 1e-4. MLP, DAE, and TCN are trained with early stopping using the Adam optimizer. The MLP model incorporates three hidden layers (64, 128, and 256 neurons) with ReLU activation and inter-layer normalization. Both DAE and TCN sub-classifiers operate on input sequences of length 300, derived from the maximum 15-second RTM action window sampled at 20 Hz. The DAE's encoder and decoder each contain three layers (encoder with 300, 128, 64 neurons and vice versa for decoder) with a 32-neuron bottleneck, utilizing MAE for reconstruction error assessment. The TCN is structured with five convolutional layers, featuring kernel size of 3, stride of 1, dilation factors [1, 2, 4, 8, 16], output channels [32, 64, 128, 256, 512], and 0.2 dropout rate. For PPO, the discount factor is set to 0.99, KL divergence limit $\beta$ to 0.02, and entropy coefficient to 0.1. An episode is defined to conclude upon reaching a maximum of 1,200 iterations or continuously violation of constraints for 10 times. As depicted in Figure \ref{converge}(a)-(c), losses for diagnostic sub-classifiers exhibit a smooth decreasing trend over epochs, stabilizing at around 400 (MLP), 300 (DAE), and 60 (TCN) iterations. Among them, the TCN demonstrates faster convergence owing to DL’s superior representation learning capability. The proximity of final training and testing losses indicates the models have not only successfully captured inherent fault patterns in the training data, but also exhibit good generalization to unseen test samples. Figure \ref{converge}(d)-(e) present the PPO agent's improving mastery of scheduling policy, with reward approaching 0 from negative values. In the starting stage (initial 100k iterations), the agent’s policy network weights are initialized randomly, signifying a limited understanding of the environment. Rapid improvements in both average reward and survival time can be observed in this stage as the agent actively explore the environment. During the transition stage (100k to 400k iterations), average rewards rise slowly while survival time dips. As the agent transitions from exploration to exploitation, it become trapped in local optima by preferentially selecting actions that were previously known to yield high rewards. In the fine-tune stage (after 400k iterations), rewards and survival time stabilize, suggesting the PPO agent has converged on optimal policies. In conclusion, the integrated diagnosis model and PPO scheduling agent achieved full convergence on the Nanjing Metro dataset without signs of overfitting or underfitting, indicating its readiness for downstream applications.

\subsection{Implementation Details of the Proposed Diagnosis Model}

\begin{figure}[h]
 \centering
 \begin{minipage}[t]{0.4\textwidth}
   \centering
   \subfloat[Incorrect segmentation with numerical approach only.]{\includegraphics[width=\linewidth]{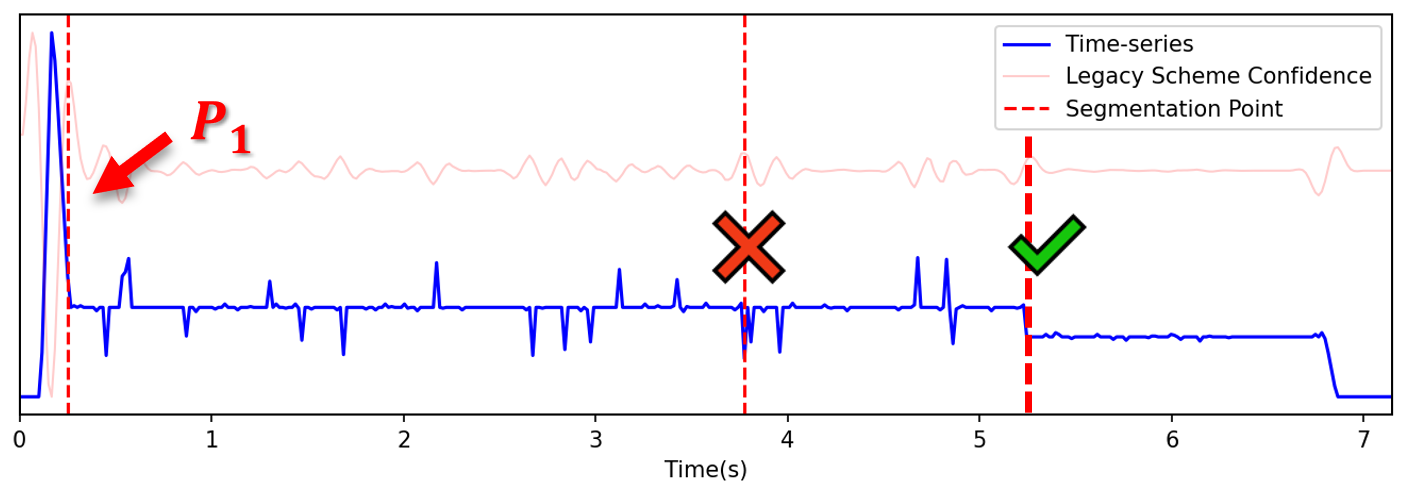}}
   \hfill
   \subfloat[Segmentation with GRU scoring mechanism.]{\includegraphics[width=\linewidth]{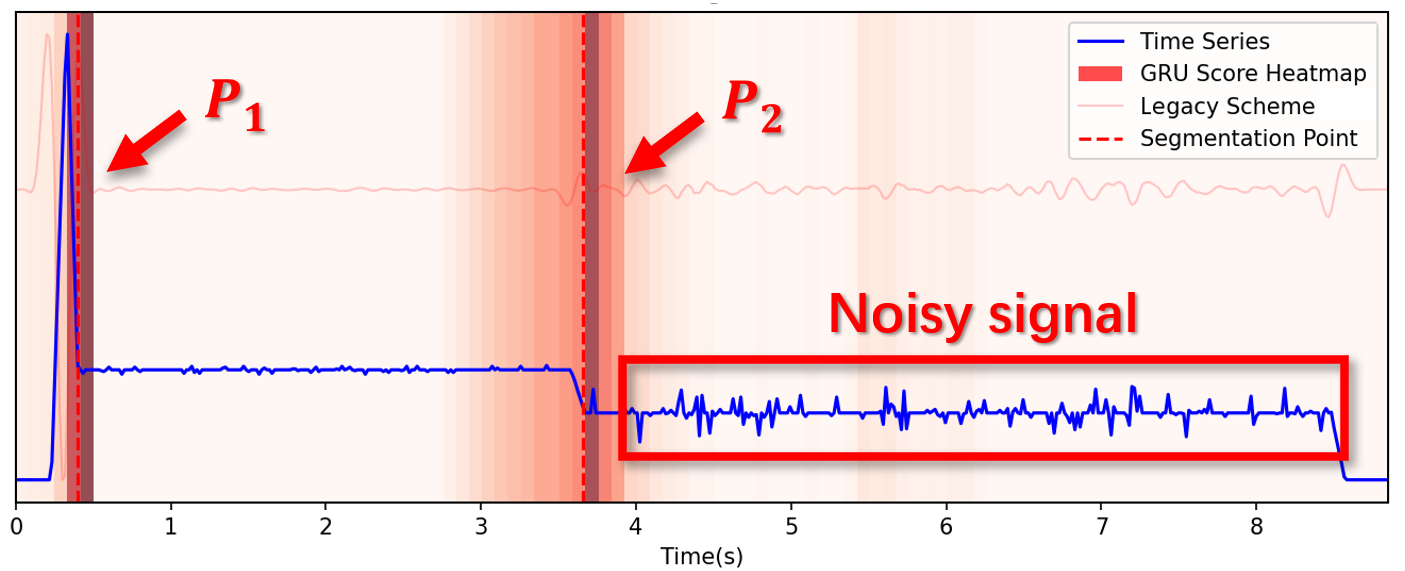}}
   \hfill
   \subfloat[Segmentation results for all sample types.]{\includegraphics[width=0.9\linewidth]{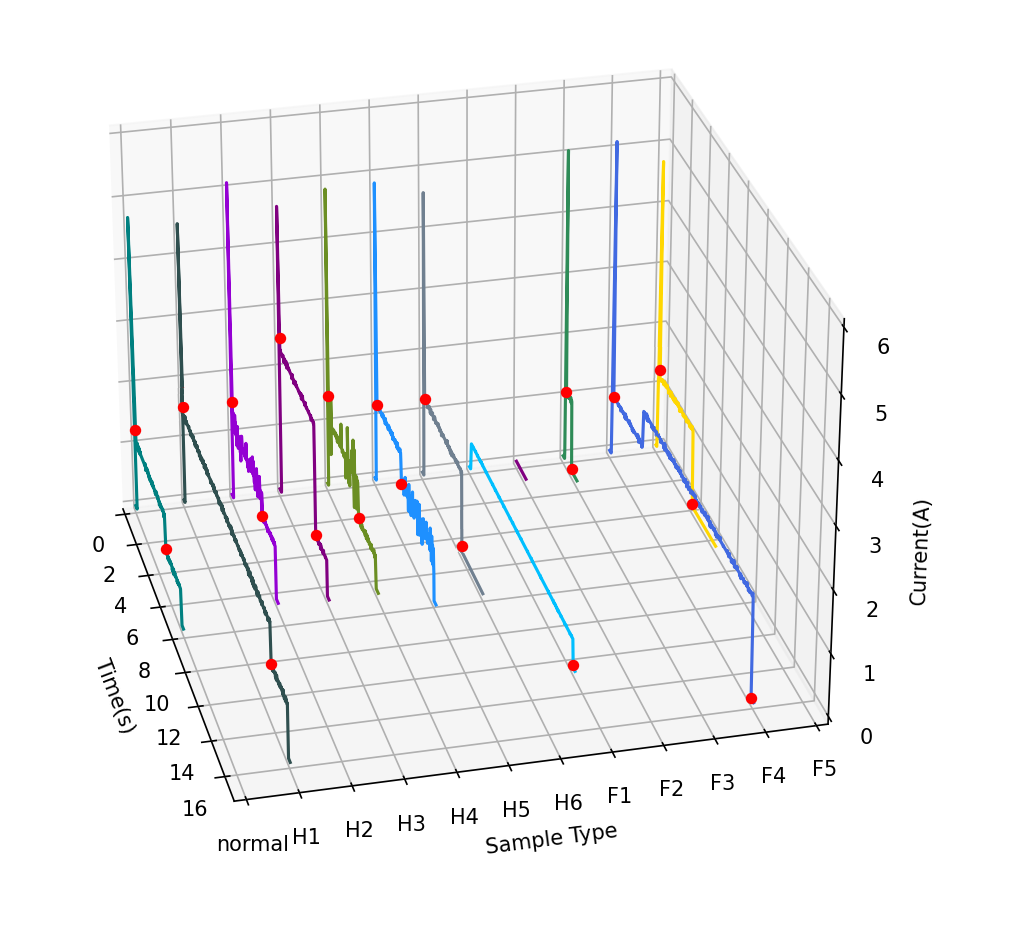}}
 \end{minipage}
 \caption{Segmentation performance demonstration with and without the proposed GRU-Based scoring technique.}
 \label{seg_result}
\end{figure}

Segmented current sequence is utilized as prior knowledge input and its accuracy directly impacts the performance of downstream classifiers. However, the standalone second-order difference-based segmentation point detection algorithm struggles dealing with fluctuating signals caused by faults and environmental electromagnetic interference. Figure \ref{seg_result}(a) shows the segmentation results for a sample with fault H4. Due to an abrupt change of current in stage 2 caused by a bad contact in the switch circuit, the numerical approach incorrectly identifies the segmentation point $P_2$. A GRU-based segmentation point confidence scoring technique is thus proposed to capture long-range dependencies in the sequence, enabling robust handling of variations and ensuring more accurate segmentation across diverse fault scenarios. Figure \ref{seg_result}(b) demonstrates the effectiveness of our refined model for a sample with fault H5 where the model correctly discerned $P_2$. The segmentation results for all fault types are shown in Figure \ref{seg_result}(c).

\begin{figure*}[b]
\vspace{-1em}
  \begin{center}
  \captionsetup[subfigure]{oneside,margin={0cm,0cm}}
    \begin{minipage}[b]{\linewidth}
        \centering
        \subfloat[MLP]{\includegraphics[width=0.25\linewidth]{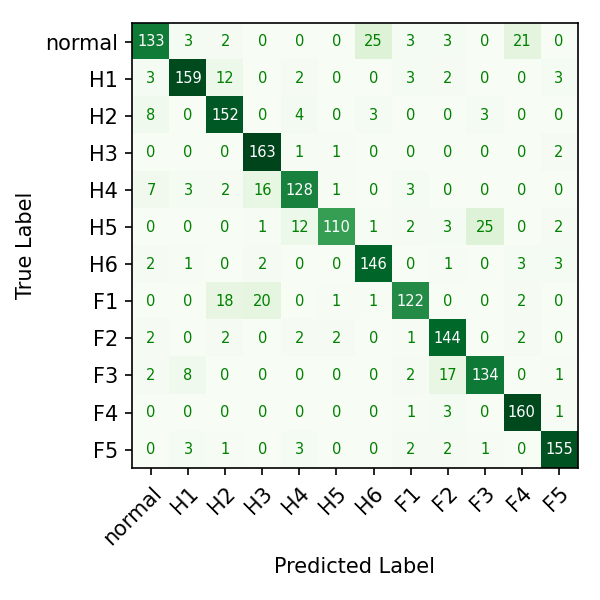}}
        \subfloat[DAE]{\includegraphics[width=0.25\linewidth]{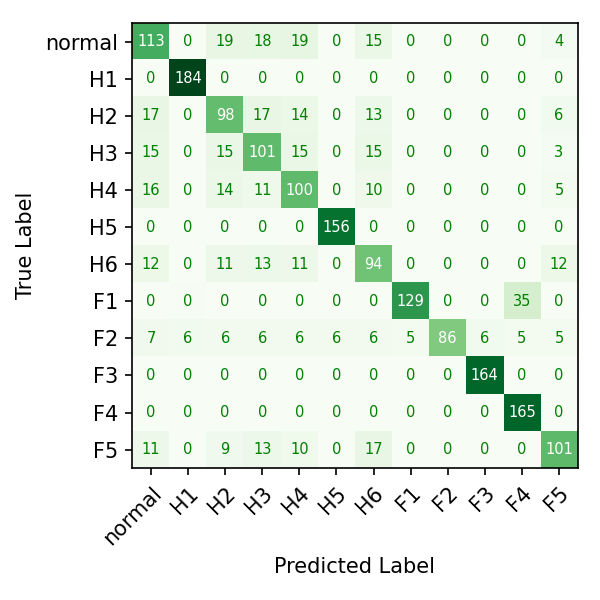}}
        \subfloat[TCN]{\includegraphics[width=0.25\linewidth]{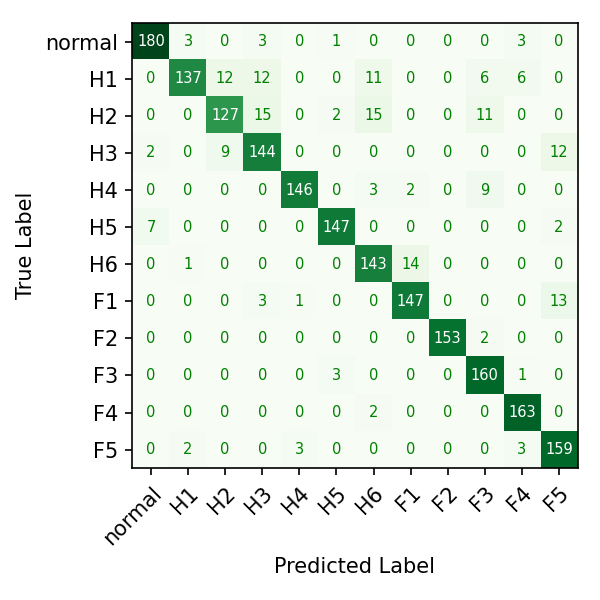}}
        \subfloat[Late-fusion output]{\includegraphics[width=0.25\linewidth]{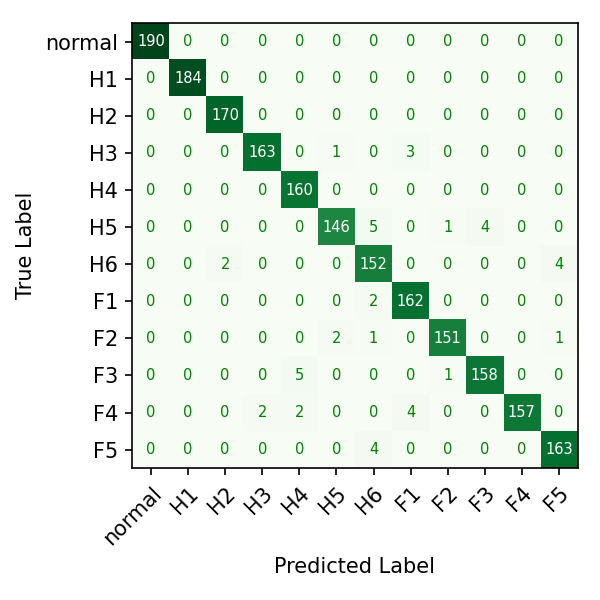}}
    \end{minipage}
  \end{center}
  \caption{Confusion matrix for sub-classifiers and their ensemble result.}
  \label{confusion_matrix}
\vspace{-1em}
\end{figure*}

DAEs are employed to discern morphological similarities among samples. The bottleneck layer feature representations for each sample type are illustrated in Figure \ref{dae_visualization}. Latent vector output in Figure \ref{dae_visualization}(a) exhibit distinct differences between fault types, indicating that DAEs have learned discriminative features during training. In Figure \ref{dae_visualization}(b), the latent vector of the normal and H1 sample types appear highly similar, while noticeable differences exist compared to types F1, F2 and F3. This suggests that the distinctiveness between latent vectors diminishes as sample morphologies become more similar. Encoding additional sample types using a DAE trained solely on normal samples can result in distorted reconstruction, which further demonstrates the feasibility of leveraging this morphological approach for fault classification.

\subsection{Ablation Study on Sub-Model Integration Results}

The proposed fault diagnosis model incorporates both prior knowledge and a sub-classifier ensemble approach. Due to limited computation resources on edge nodes, all sub-classifiers are uniformly crafted with shallow network architectures, potentially limiting their ability to capture abstract and complex patterns from the input sequence. While the introduction of prior knowledge can enhance accuracy, it also incurs additional computational overhead. Therefore, an ablation study is conducted by formulating four model variants that systematically remove model components. As shown in Table \ref{ablation_study}, this experimental setup enables assessment of each component's contribution to the overall classification performance and inference speed. Notably, parallel optimization is disabled for this set of experiments and serial execution on a single machine is enforced between all sub-models to facilitate data analysis.

\begin{table}[h]
\caption{Ablation study results. ``AW'' means the amount of the weights and ``AT'' means the average response time.}
\label{ablation_study}
\centering
\renewcommand{\arraystretch}{1.4}
\begin{adjustbox}{width=\columnwidth,center}
\begin{tabular}{c|ccccc|cc}
\hline
\multirow{2}{*}{\textbf{Methodology}} & \multicolumn{5}{c|}{\textbf{Classification Performance}} & \multicolumn{2}{c}{\textbf{Inference Speed}} \\
\cline{2-6,7-8}
                                    & \textbf{Accu} & \textbf{Prec} & \textbf{Recall} & \textbf{F1} & \textbf{FPR} & \textbf{AW} & \textbf{AT (ms)}    \\
\hline
w/o Prior                      & 0.612   & 0.583   & 0.697    & 0.635   & 0.324   & 1.86M            & \textbf{421.84}             \\
w/o MLP                        & 0.870   & 0.757   & 0.821    & 0.788   & 0.159   & 2.01M            & 849.40             \\
w/o DAE                        & 0.914   & 0.881   & 0.903    & 0.892   & 0.083   & 1.62M            & 744.58             \\
w/o TCN                        & 0.781   & 0.712   & 0.750    & 0.731   & 0.229   & \textbf{0.82M}             & 653.19             \\
Original                           & \textbf{0.974} & \textbf{0.969} & \textbf{0.991} & \textbf{0.980} & \textbf{0.013}   & 2.08M            & 889.67 \\\hline
\end{tabular}
\end{adjustbox}
\end{table}

TCN stands out as the most robust classifier, making the most substantial contribution to the final results. This is evident from the fact that its removal results in the most dramatic decreases in both False Positive Rate (FPR) and recall rate among the four model variants. Clear trade-offs can be observed between classification performance and computational requirements. The model without prior knowledge achieves the largest inference time reduction of 467.83ms. This improvement can be attributed to the elimination of sequential GRU cell processing, as well as simplifying the downstream workflow without stage-wise processing. Additionally, excluding TCN yields the greatest reduction in model parameters due to its deeper architecture. MLP demonstrates the fastest prediction time, owing to its feature extraction process being primarily based on numerical computations. Notably, the original model which incorporates all components, achieves the highest classification performance across all evaluation metrics. Besides, the performance gain corresponding to the increase in computational load is considerable. This observation highlights the effectiveness of the ensemble approach in stacking sub-models.

Detailed analysis of the sub-classifiers' performance across each fault type is presented in Figure \ref{confusion_matrix}. Specifically, Figure \ref{confusion_matrix}(b) for the DAE sub-classifier exhibits some mutual misclassification among H2, H3, and H4 type of faults, which is due to these faults being morphologically similar after reconstruction. Consistent with our earlier observations, TCN (Figure \ref{confusion_matrix}(c)) performs the best among all sub-classifiers with most of its predictions concentrated along the diagonal, indicating fewer errors across classes. The performance of the proposed FL fusion scheme, which combines outputs from MLP, DAE, and TCN to form the final prediction, is illustrated in Figure \ref{confusion_matrix}(d). We observe a higher prevalence of correct classifications throughout the FL fusion scheme's confusion matrix, signifying its enhanced ability to produce accurate predictions across all fault types. These findings suggest that our stacked model has developed a nuanced and comprehensive perception of fault patterns, leading to a fault detection system that is robust, flexible and future-proof.

\begin{table*}[b]
\caption{Comparison of Scheduling Schemes Under Different Workloads}
\label{comparison}
\centering
\renewcommand{\arraystretch}{1.2}
\begin{threeparttable}
\begin{tabular}{ccccccc} 
\hline
\multirow{2}{*}{\textbf{Request Rate}} & \multirow{2}{*}{\textbf{Scheduling Scheme}} & \multicolumn{3}{c}{\textbf{Avg Response Time (ms)}} & \multirow{2}{*}{\begin{tabular}[c]{@{}c@{}}\textbf{Load Balancing\tnote{a)}}\\\textbf{(weighted std)}\end{tabular}} & \multirow{2}{*}{\textbf{Resource Utilization}} \\ 
\cline{3-5}
& & \textbf{Computation} & \textbf{Transmission} & \textbf{Total} & & \\
\hline
\multirow{5}{*}{\begin{tabular}[c]{@{}c@{}}10 req/s \\(low workload)\end{tabular}} 
& Random & 736.56 & 16.35 & 752.91 & 23.36 & 13.35 \\
& Round Robin & 816.09 & 14.49 & 830.58 & 27.63 & 15.92 \\
& EPS & 866 & 23.04 & 889.04 & 28.25 & 12.01 \\
& CPS & \textbf{486.86} & 14.50 & \textbf{501.36} & 47.03 & 16.34 \\
& \textbf{Ours} & 495.84 & \textbf{13.92} & 509.76 & \textbf{20.12} & \textbf{16.49} \\ 
\hline
\multirow{5}{*}{\begin{tabular}[c]{@{}c@{}}50 req/s \\(medium workload)\end{tabular}} 
& Random & 878.7 & 24.51 & 903.21 & 25.21 & 65.08 \\
& Round Robin & 828.62 & 20.45 & 849.07 & 23.57 & 61.31 \\
& EPS & 1179.88 & 26.27 & 1206.15 & 21.21 & 59.08 \\
& CPS & 752.06 & 17.38 & 769.44 & 33.43 & 68.36 \\
& \textbf{Ours} & \textbf{621.61} & \textbf{16.19} & \textbf{637.80} & \textbf{17.66} & \textbf{82.12} \\ 
\hline
\multirow{5}{*}{\begin{tabular}[c]{@{}c@{}}200 req/s \\(high workload)\end{tabular}}  
& Random & 9098.77 & 102.21 & 9200.98 & 19.11 & 71.30 \\
& Round Robin & 8913.2 & 131.84 & 9045.04 & 18.04 & 70.54 \\
& EPS & 9445.04 & 357.67 & 9802.71 & 18.62 & 63.95 \\
& CPS & 4812.17 & 63.07 & 4875.24 & 23.18 & 74.02 \\
& \textbf{Ours} & \textbf{1662.78} & \textbf{46.56} & \textbf{1709.34} & \textbf{13.28} & \textbf{99.57} \\
\hline
\end{tabular}
    \begin{tablenotes}
     \item[a)] The load balancing metric evaluates workload distribution in accordance with the relative computational capabilities of each node, enabling a more precise evaluation of system efficiency. Let $\sigma_w$ represent the weighted standard deviation, $W_i$ denote the weight of the i-th node (proportionate to its computational capacity within the system), $x_i$ signify the workload of the i-th node, and $\bar{x}_w$ denote weighted mean of the workloads. This metric can be calculated as $\sqrt{{\sum_{i=1}^{n} W_i (x_i - \bar{x}_w)^2}/{\sum_{i=1}^{n} W_i}}$.
   \end{tablenotes}
\end{threeparttable}
\end{table*}

\subsection{Performance Analysis of CEC-PA Under Network Degradation}

In railway transportation scenarios, network communication quality is significantly affected by multiple factors, including dense user devices interference, Doppler effects from high-speed train movement, and signal attenuation in underground tunnels. The proposed CEC-PA framework addresses these challenges through DRL-based adaptive task offloading and implements a consensus-based coordinator node election mechanism to maintain system robustness during node downtimes. To evaluate CEC-PA's performance under weak network conditions, experiments are conducted under a typical request frequency of 50 req/s with no request timeout limit. Additional delays were probabilistically introduced across 100 recent D2D data transmissions to simulate the impact of varying latency and packet loss. Results are presented in Table \ref{critical_network_table}, where the horizontal and vertical axes represent the percentage of affected connections and different levels of added network delay respectively, with each cell indicating the average response time for each condition combination.

Under optimal network conditions (no added delay), the system maintains a baseline response time of 637.80ms, demonstrating CEC-PA's robust performance in near-ideal scenarios. Even when 100\% of connections experience with mild network degradation (20ms delay added), the response time increases by only 27.1\% to 810.63ms. This moderate impact is attributed to CEC-PA's adaptive task scheduling mechanism, which effectively redistributes workload to compensate for network perturbations. As network delay increases to 100ms, the system starts to demonstrate non-linear performance degradation. When 100\% of connections are affected, the response time increases by 140\% to 1529.71ms. This more pronounced impact reflects the cumulative effect of greater reliance on multi-hop communication paths, task synchronization and increased frequency of retransmission attempts. Under severe network stress (500ms delay added), the system experiences significant performance impact, with response times increasing by 660\% to 4850.42ms when 100\% of connections are affected. However, when 20\% of connections experience such delays, the impact is contained to a 78.4\% increase (1137.68ms), demonstrating the system's partial resilience through connection diversity. During packet loss ($\infty$ delay added) conditions, the system maintains operability until complete network failure, with response times reaching 26010.42ms at 80\% of connection affected.

\begin{table}[h]
\caption{Average Response Time (ms) Under Different Network Conditions}
\label{critical_network_table}
\centering
\renewcommand{\arraystretch}{1.2}
\begin{tabular}{cccccc}
\hline
\multirow{2}{*}{\textbf{Delay Added}} & \multicolumn{5}{c}{\textbf{Connections   Affected}}     \\   \cline{2-6}
                             & \textbf{20\%}    & \textbf{40\%}    & \textbf{60\%}     & \textbf{80\%}     & \textbf{100\%}   \\ \hline
+0ms                         & \multicolumn{5}{c}{637.80}                \\ 
+20ms                        & 652.74  & 669.85  & 682.49   & 739.96   & 810.63  \\
+100ms                       & 837.89  & 879.63  & 915.04   & 1249.57  & 1529.71 \\
+500ms                       & 1137.68 & 1358.34 & 1992.97  & 3076.64  & 4850.42 \\
\multirow{2}{*}{\begin{tabular}[c]{@{}c@{}} +$\infty$ \\ (Packet Loss) \end{tabular}}             & \multirow{2}{*}{\centering 3172.81} & \multirow{2}{*}{\centering 7653.48} & \multirow{2}{*}{\centering 12436.15} & \multirow{2}{*}{\centering 26010.42} & \multirow{2}{*}{N/A}  \\ \\
\hline
\end{tabular}
\end{table}

These findings validate CEC-PA's resilience in maintaining acceptable performance under varying network conditions, with graceful degradation of service quality rather than catastrophic failures. The DRL scheduler with downtime-tolerance mechanisms effectively prevent system offline even under severe network impairment, ensuring continuous operation of the fault diagnosis system.

\subsection{Runtime Performance Comparison with Existing Scheduling Schemes}

In section \ref{section_cec}, our proposed DRL-based scheduling framework CEC-PA is designed to perform optimal decision-making in real-time, dynamically adapting to the complex and dynamic state inputs from the distributed environment. To showcase the effectiveness of CEC-PA, four classic baselines are selected for comparative experimentation, including:

\begin{itemize}
    \item \textbf{Random:} Tasks are randomly assigned to edge or cloud nodes.
    \item \textbf{Round Robin:} Nodes take turns receiving tasks in a fixed sequential order.
    \item \textbf{Edge-preference Scheduling (EPS):} Prioritizes assigning tasks to the edge, offloading to the cloud only when necessary.
    \item \textbf{Cloud-preference Scheduling (CPS):} Prioritizes assigning tasks to the cloud, opting for the edge only when cloud resources are fully occupied.
\end{itemize}

Real-world workloads are typically volatile and unpredictable, and their impact on scheduling decisions should not be overlooked. Experiments are conducted to evaluate how these scheduling schemes perform under different workloads, simulating request rates at 10 req/s, 50 req/s, and 200 req/s. The results are presented in Table \ref{comparison}, where load balancing is measured by the standard deviation of workload distribution across nodes weighted by their computing capacities and resource utilization is the percentage of the computation power in use.

\begin{table*}[b]
\caption{Comparative Analysis of RTM Fault Diagnosis Schemes}
\label{overall_comp}
\centering
\renewcommand{\arraystretch}{1.2}
\begin{tabular}{cccccc|ccc}
\hline
\multirow{2}{*}{\textbf{Methodology}} & \multicolumn{5}{c|}{\textbf{Classification   Performance}}                           & \multicolumn{3}{c}{\textbf{Avg Response Time (ms)}}        \\  \cline{2-9}
                                      & \textbf{Accu}  & \textbf{Prec}  & \textbf{Recall} & \textbf{F1}    & \textbf{FPR}   & \textbf{10 req/s} & \textbf{50 req/s} & \textbf{200 req/s} \\  \hline
SVM                                   & 0.690          & 0.877          & 0.593           & 0.707          & 0.143          & \textbf{104.29}   & 659.06            & 2301.67            \\
GBDT                                  & 0.842          & 0.936          & 0.806           & 0.866          & 0.095          & 252.69            & 921.04            & 4134.72            \\
DNN                                   & 0.912          & 0.919          & 0.944           & 0.931          & 0.143          & 395.12            & 2195.27           & 9288.06            \\
EBTW+1DCNN                     & 0.924          & 0.899          & \textbf{0.991}  & 0.943          & 0.191          & 371.42            & 1683.75           & 8229.87            \\
AE+GRU                         & 0.942          & 0.971          & 0.935           & 0.953          & 0.048          & 796.31            & 4672.39           & 22539.18           \\
Ours                                  & \textbf{0.974} & \textbf{0.969} & \textbf{0.991}  & \textbf{0.980} & \textbf{0.013} & 509.76            & \textbf{637.80}   & \textbf{1709.34}  \\ \hline

\end{tabular}
\end{table*}

In scenarios with low workloads, both CPS and CEC-PA demonstrate the shortest total response time of around 500 milliseconds. This is because cloud resources are abundant and requests can be handled without necessitating the involvement of edge nodes. Edge nodes possess lower individual computational power but are distributed in greater numbers. However, with sparse workloads, the edge nodes remain underutilized and are unable to manifest their capabilities. This results in the edge-centric EPS exhibiting the longest response time of 889.04 milliseconds. In medium workload scenarios, computation time of the edge-centric EPS experiences a significant increase by 36.14\%, while there are no notable changes in its transmission time. This suggests that the edge nodes start to reach capacity bottlenecks, leading to longer queueing delays for task partitions. We observe no major differences in resource utilization rates across different scheduling schemes under low workload. However, divergences emerge under medium workload and above. CEC-PA achieves the highest resource utilization rate of 82.12\% to 99.57\% compared to others by dynamically leveraging both edge and cloud nodes to avoid over-reliance on either type of these resources. During periods of high concurrency, scheduling schemes including Random, Round Robin and EPS experience numerous timeouts due to wide-spread overloading of nodes, thus causing a sharp increase in the average response time.

Under various load conditions, CEC-PA consistently outperforms static baselines in terms of both resource utilization and response time. During peak loads on edge nodes, CEC-PA dynamically shifts more tasks to the cloud, avoiding potential timeouts caused by queuing delays. Meanwhile, when CEC-PA detects low cloud resource utilization or decreased edge node loads, it adjusts its strategy by increasing the proportion of tasks allocated to edge nodes. Performance metrics for all scheduling schemes worsen dramatically in high workload scenarios, except for CEC-PA, which continues to maintain good performance. Under the highest simulated workload of 200 req/s, CEC-PA improves response time by 280\% and boosts resource utilization by 35\% compared to the next best scheme. The experiment results presented in Table \ref{comparison} validate the effectiveness of CEC-PA's adaptive scheduling strategy that conducts intelligent decision-making based on real-time node conditions.

\subsection{Comparative Analysis of CEC-PA’s Pipeline Partitioning Scheme}
\begin{figure}[h]
\centering
  \begin{minipage}[b]{0.85\linewidth}
    \centering
    \subfloat[Computation time]{\includegraphics[width=\linewidth]{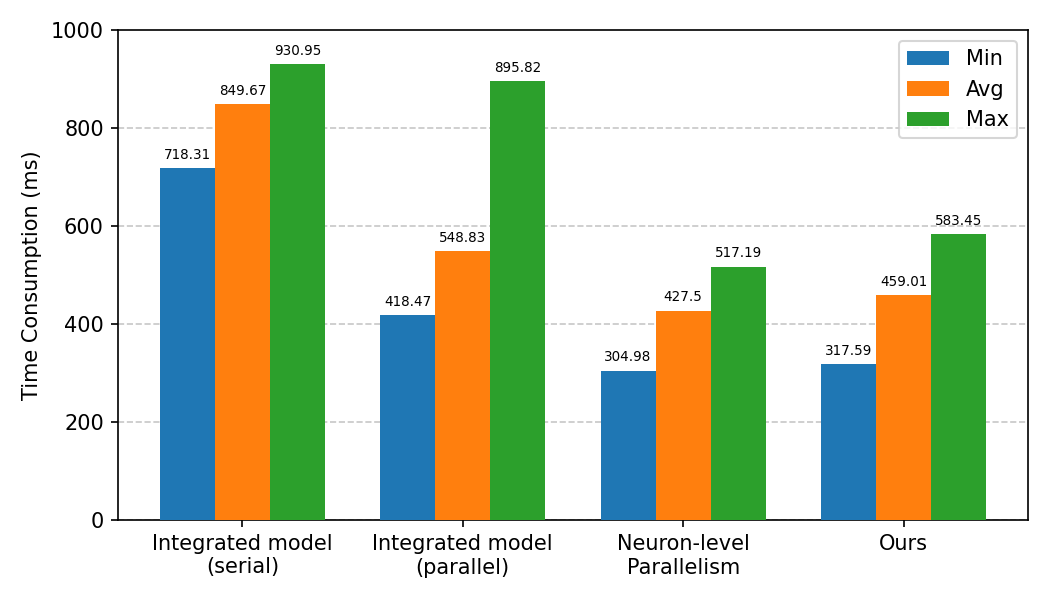}}
    \hfill
    \subfloat[Transmission time]{\includegraphics[width=\linewidth]{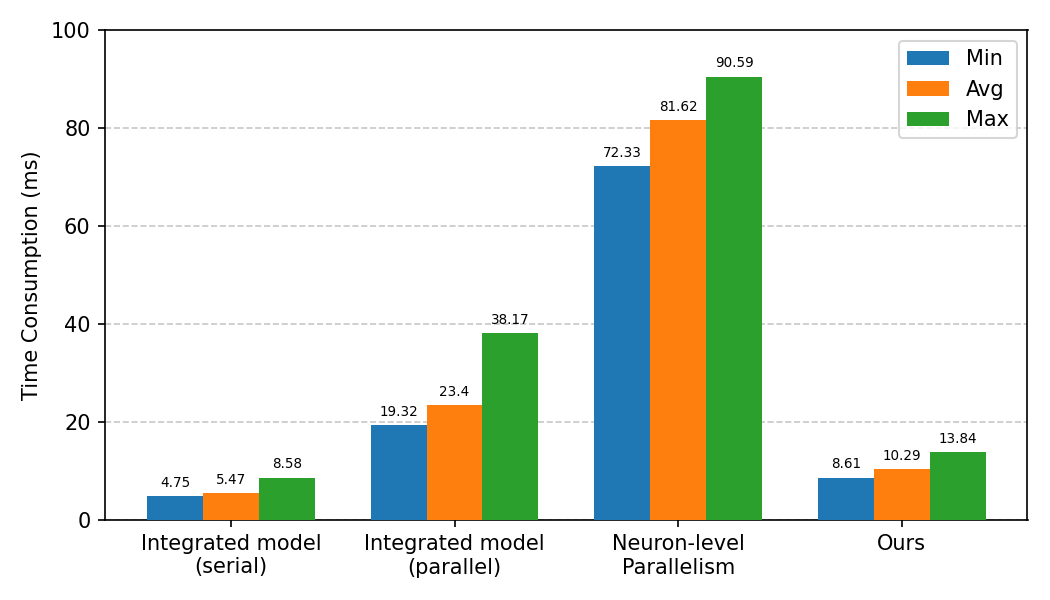}}
  \end{minipage}
  \caption{Comparison on different partitioning granularity in conjunction with various execution strategies.}
  \label{bar_chart}
  \vspace{-1em}
\end{figure}

In Section \ref{section_6.1}, we proposed a partitioning scheme within our CEC-PA framework for parallelism optimization of model components. To evaluate the effectiveness of the proposed partitioning scheme, a comparison of different partition schemes is conducted. For the control group of model components without partitioning, we considered two execution paradigms as baselines: full-serial and full-parallel. In the full-serial paradigm, model components are executed in a strict order which eliminates parallelism. Conversely, the full-parallel execution paradigm maximizes parallelism by directly assigning all model components to the scheduler. Additionally, neuron-level parallelism \cite{10177476} is selected for comparison due to its potential to achieve the highest degree of parallel processing within the neural network scope. The experimental results are shown in Figure \ref{bar_chart}.

The comparison between full-serial and full-parallel execution paradigms reveals the trade-offs between sequential simplicity and parallel efficiency. The full-serial paradigm keeps only one model component active at each time slot, with its output directly propagated to the next component without checking the completion status of other worker nodes. This results in relatively low communication overhead of approximately 50\% between nodes. Although the full-serial approach demonstrates excellent context transmission time, its computational time strikes the highest due to its lack of parallelism. In contrast, the full-parallel approach significantly reduces computation time by simultaneously executing all components. However, unnecessary communication overheads emerge from frequent data exchanges between concurrently active components, leading to slightly higher transition time compared to full-serial approach and CEC-PA. The issue of transmission overhead becomes even more pronounced in neuron-level parallelism, which offers parallelism at the finest granularity within neural network layers. While achieving the lowest computation time of 477.5 milliseconds in average, it also results in 7.93x higher transmission time than CEC-PA due to the intricate data exchanges required during model weights propagation.

In conclusion, the proposed CEC-PA partitioning scheme demonstrates superior performance against other paradigms when considering both computation and transmission overhead. By packing model components based on their resource requirement similarity and contextual dependencies, it strikes a balance between data exchange and parallelism. Quantitatively, it achieves up to 1.98x computation speed-up over full-serial approach and 7.93x transmission speed-up over neuron-level parallelism. Such strategic partitioning scheme paired with its coordinated pipeline scheduling policy establishes an efficient and streamlined computational framework ideal for the distributed turnout fault detection.

\subsection{Overall Comparison with Well-established RTM Fault Diagnosis Schemes}

To comprehensively validate the efficacy of our proposed approach, we conducted an extensive comparative analysis against several well-established RTM fault diagnosis schemes. Each baseline was carefully reproduced and evaluated using our dataset, with comparative performance metrics presented in Table \ref{overall_comp}.

Achieving an accuracy of 97.4\%, the proposed scheme outperforms conventional SVM (69.0\%) and GBDT (84.2\%) approaches by margins of 28.4\% and 13.2\%, respectively. Notably, the system achieves a remarkably low false positive rate of 0.013, marking a 72.9\% reduction relative to the next-best performing AE+GRU scheme. These significant performance gains can be primarily attributed to our novel integration of domain knowledge-driven feature extraction with advanced DL stacking architecture for pattern recognition. In terms of inference speed, the proposed scheme exhibits remarkable scalability under increasing workloads. While performing competitively at low request rates (509.76ms average response time at 10 req/s), it demonstrates exceptional efficiency at higher load scenarios. The request rate from 10 to 200 req/s shows only a 3.4x increase in inference time, contrasting sharply with the 28.3x increase observed in the AE+GRU scheme. This superior scalability stems from our optimized pipeline partitioning strategy and cloud-edge collaborative framework, which effectively distributes computational loads and minimizes communication overhead. 

The dramatic improvements in both classification performance and inference speed suggest that our approach successfully addresses the traditional trade-off between classification accuracy, model complexity and responsiveness. This is particularly evident in high-load scenarios where competing methods exhibit significant performance degradation. The proposed scheme maintains its responsiveness and accuracy even under harsh conditions, underlining its robustness and adaptability to varying operational demands.

\section{Conclusion and Future Work}
\label{conclusion}
As a critical safety measure, the turnout fault early-warning system needs to deliver timely and accurate diagnostic results on a continuous 7x24 basis. This research aims to address the real-time and robustness challenges of turnout fault diagnosis systems through an edge-cloud collaborative deployment approach. Specifically, a parallel-optimized fault classification model with ensemble technique and prior knowledge is proposed. Then, the integrated model is further partitioned into pipelines and scheduled across edge and cloud via the CEC-PA framework, which enables efficient and flexible computation offloading. Although the experimental results demonstrate promising outcomes, there still remain several avenues for future enhancement. One potential direction is to optimize the MDP modeling to further improve the system's decision-making capabilities. Besides, a backup node election consensus mechanism can be proposed to ensure uninterrupted operation of the coordinator node in cloud downtime.

\section*{Acknowledgment}
This work was supported by the National Natural Science Foundation of China under Grant (No. 62372242 and 92267104), and in part by Natural Science Foundation of Jiangsu Province of China under Grant (No. BK20211284).


%





\ifCLASSOPTIONcaptionsoff
  \newpage
\fi





\bibliographystyle{IEEEtran}
\bibliography{IEEEabrv,Bibliography}
%

\begin{IEEEbiography}[{\includegraphics[width=1in,height=1.25in,clip,keepaspectratio]{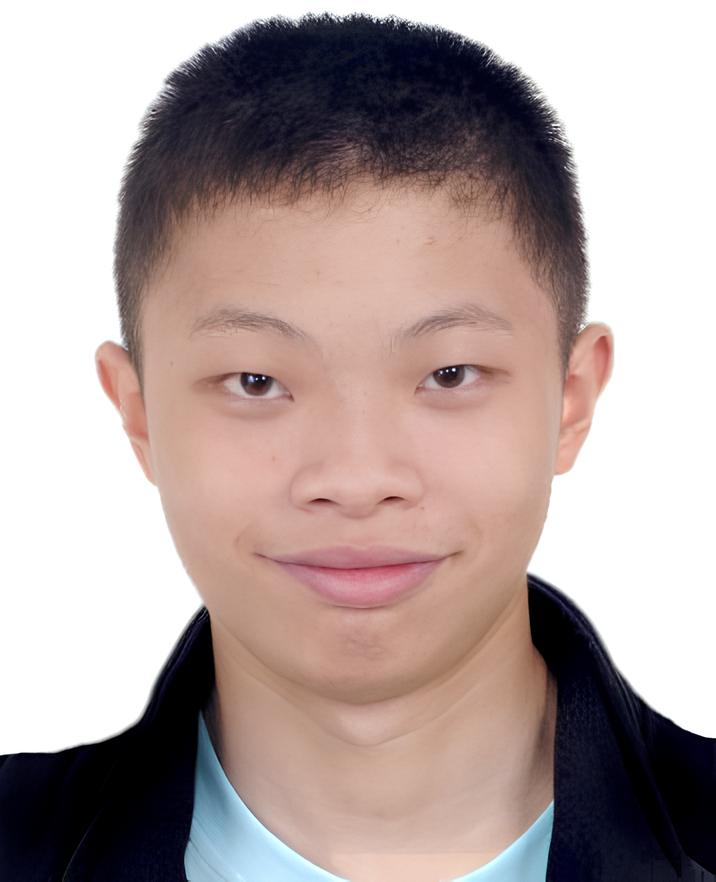}}]{Fan Wu}
is currently a postgraduate student at the School of Software, Nanjing University of Information Science and Technology, China. His research interests include mobile edge computing, fault diagnosis, etc.
\end{IEEEbiography}

\begin{IEEEbiography}
[{\includegraphics[width=1in,height=1.25in,clip,keepaspectratio]{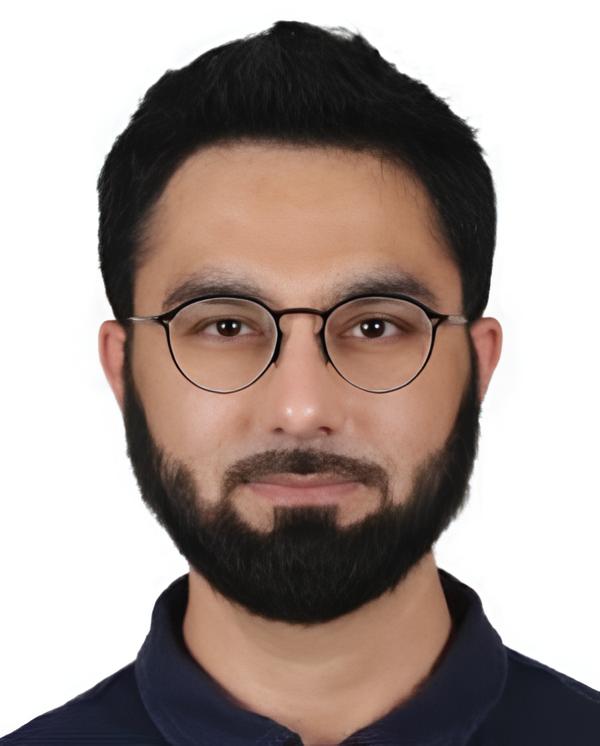}}]{Muhammad Bilal}
received the Ph.D. degree in information and communication network engineering from the School of Electronics and Telecommunications Research Institute (ETRI), Korea University of Science and Technology, Daejeon, South Korea, in 2017. From 2017 to 2018, he was with Korea University, where he was a Postdoctoral Research Fellow with the Smart Quantum Communication Center. In 2018, he joined the Hankuk University of Foreign Studies, South Korea, where he was an Associate Professor with the Division of Computer and Electronic Systems Engineering. He is currently a Senior Lecturer (Associate Professor) with the School of Computing and Communications, Lancaster University, Lancaster, U.K.
\end{IEEEbiography}

\begin{IEEEbiography}[{\includegraphics[width=1in,height=1.25in,clip,keepaspectratio]{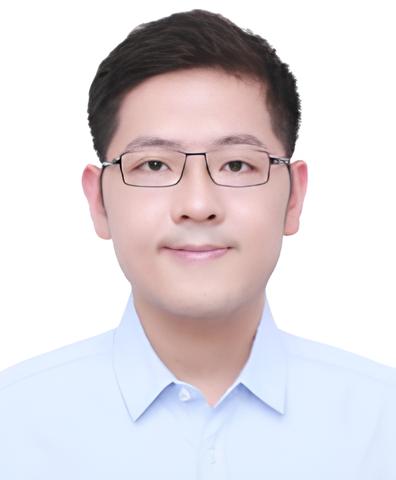}}]{Xiaolong Xu}
received the Ph.D. degree in computer science and technology from Nanjing University, China, in 2016. He is currently a Full Professor with the School of Software, Nanjing University of Information Science and Technology. He received the Best Paper Awards from the IEEE CBD 2016, IEEE CyberTech2021, IEEE iThings2022 and IEEE ISPA 2022, and the Outstanding Paper Award from IEEE SmartCity2021. He received the Outstanding Leadership Award of IEEE UIC 2022. He also received the Best Paper Award from Elsevier JNCA. He has been selected as the Highly Cited Researcher of Clarivate 2021 and 2022. His research interests include edge intelligence and service computing.
\end{IEEEbiography}





\vfill


\end{document}